\documentclass[pdftex,nofootinbib,superscriptaddress,aps,prfluids,notitlepage,longbibliography]{revtex4}
\usepackage{multirow,enumerate,epsfig,graphics,amssymb,amsmath,subeqnarray,mathrsfs,color,xcolor,epstopdf}
\usepackage{natbib}
\usepackage{color,epsfig,graphics,amssymb,mathcmd,amsmath,subeqnarray,graphicx,amsthm,subfigure,mathrsfs}
\usepackage[colorlinks=true, citecolor=red, linkcolor=blue ]{hyperref}
\usepackage{mathtools,bm} 
\usepackage{rotating,mathrsfs}
\usepackage[normalem]{ulem}

\makeatletter
\def\sgn{\mathop{\operator@font sgn}}
\def\threevdots{\vbox{\baselineskip1\p@ \lineskiplimit\z@
  \kern6\p@\hbox{.}\hbox{.}\hbox{.}}}
\makeatother

\begin{document} 

\title{Hydro-chemical interactions in dilute phoretic suspensions: from individual particle properties to collective organization}
\author{T. Traverso}
\email{traverso@ladhyx.polytechnique.fr}
\affiliation{LadHyX -- D\'epartement de M\'ecanique, Ecole Polytechnique -- CNRS, 91128 Palaiseau Cedex, France}
\author{S. Michelin}
\email{sebastien.michelin@ladhyx.polytechnique.fr}
\affiliation{LadHyX -- D\'epartement de M\'ecanique, Ecole Polytechnique -- CNRS, 91128 Palaiseau Cedex, France}

\begin{abstract}
Janus phoretic colloids (JPs) self-propel as a result of self-generated chemical
gradients and exhibit spontaneous nontrivial dynamics within phoretic suspensions,
on length scales much larger than the microscopic swimmer size. Such collective
dynamics arise from the competition of (i) the self-propulsion velocity of the particles,
(ii) the attractive/repulsive chemically-mediated interactions between particles and (iii)
the flow disturbance they introduce in the surrounding medium. These three
ingredients are directly determined by the shape and physico-chemical properties of
the colloids' surface. Owing to such link, we adapt a recent and popular kinetic model for dilute
suspensions of chemically-active JPs where the particles' far-field hydrodynamic and
chemical signatures are intrinsically linked and explicitly determined by the design
properties.
Using linear stability analysis, we show that
self-propulsion can induce a wave-selective mechanism for certain particles'
configurations consistent with experimental observations. Numerical simulations of
the complete kinetic model are further performed to analyze the relative importance
of chemical and hydrodynamic interactions in the nonlinear dynamics. Our results
show that regular patterns in the particle density are promoted by chemical signaling
but prevented by the strong fluid flows generated collectively by the
polarized particles, regardless of their chemotactic or
antichemotactic nature (i.e. for both puller and pusher swimmers).
\end{abstract}
\maketitle
\section{Introduction}

The complex self-organization and collective dynamics of microswimmers within so-called active suspensions have fascinated many researchers across disciplines, in part because of their ubiquity in the biological world but also as a simple model system to study the emergence of collective behavior. 
As an archetypal example of matter out of equilibrium displaying  rich phenomenology, such complex systems have been studied and described with methods akin to those widely employed in statistical mechanics and thermodynamics ~\cite{Marchetti13,Marconi2019_SoftMatter,Role_of_Correlations_Morozov2017}.
The collective motion of simple living microorganisms (e.g. bacteria) as well as their capacity to self-organize~\cite{Dombrowski2004} and to respond to external stimuli~\cite{Rafai2016_rapidPRE} have also inspired the design of artificial micro-scale swimmers, including for potential engineering applications such as cargo transport for targeted drug delivery \cite{Popescu2011_Akhil_SM_21,Akhil_SM_20_Wang2013} or as micromachines \cite{Catchmark2005_Akhil_SM_22,sokolov2010}. 

The ability of such artificial micro-scale swimmers to self-propel through the surrounding viscous medium is essential to the emergence of non-trivial collective dynamics or for technological applications, and can be forced externally using acoustic vibrations \cite{Vibrartions_nature_Ahmed} or electro-magnetic fields~\cite{Ghosh09,Debasish_PRL2019_electricField,Blaise_microrollersMagnetic_NatureP}.
Alternatively, direct interactions of individual particles with their immediate environment can convert physico-chemical energy to set the fluid  into motion and self-propel~\cite{moran2017,ebbens2016}. To this end, the particles' surface must possess two fundamental physico-chemical properties: (i) a phoretic \emph{mobility} to convert  physico-chemical gradients into  mechanical forcing locally~\cite{anderson89} and (ii) an \emph{activity} to create such local gradients without any external directional forcing~\cite{Duan15}. This activity, which can take diverse forms such as chemical catalysis or heat release, can be either spontaneous (e.g. hydrogen peroxyde decomposition on gold-platinum colloids~\cite{Theurkauff2012,Ginot2018}) 
or externally-activated (e.g. via an optical field \cite{palacci2013living,Buttinoni2013}), providing a route to remote control of the activity of the colloids and of their self-propulsion without direct imposition of a mechanical forcing. 

Directed self-propulsion at the microscopic scale also requires breaking the spatial symmetry of the mechanical forcing, and thus of the physico-chemical environment of the particles, which is most commonly achieved by an asymmetric design of the particles (e.g. surface coating) as for Janus phoretic colloids with two distinctly-coated ends~\cite{ebbens2016,moran2017}. 
On the other hand, motile microorganisms 
use motile appendices (cilia or flagella) whose coordinated movement produces non-reciprocal waves that break the time-reversal symmetry and guarantees propulsion at the microscopic scale \cite{pak2014theoretical,Stone1996}. Despite such difference in how microorganisms and JPs achieve sustained directional motion, the slowest decaying mode of the flow disturbance generated by both kind of swimmers is the one of a force-free self-propelling particle in a viscous medium, namely a force-dipole (or stresslet) \cite{batchelor_1970,pak2014theoretical,blake_1971,Michelin2014}. Such far-field hydrodynamic signature 
dominates the hydrodynamic interactions between swimmers within dilute suspensions and it is the only retained contribution in far-field models. The hydrodynamic coupling of JPs as well as of microorganisms is therefore modeled in similar ways. 

The ability to generate gradients of the physico-chemical properties of the surrounding medium and to respond to them is another shared feature between JPs and certain microorganisms (e.g. \textit{Escherichia Coli}) and allows swimmers to interact chemically within the suspension. While the analogy holds from a phenomenological point of view, it can not be extended to the details of the physical mechanisms governing the attractive (or repulsive) nature of the chemically mediated interactions.  
Specifically, by virtue of the phoretic mechanism the mere presence of a chemical gradient, which can result from the presence of nearby particles, set Janus particles into motion along its direction, a distinguished feature of phoretic colloids \cite{Golestanian2007,palacci2013living}. Alternatively, autophoretic particles with non-uniform surface mobility reorient along an external gradient under the effect of the induced aligning torque and self-propel in such direction \cite{Tatulea-Codrean2018,Saha2014,Kanso2019}. On the other hand, microorganisms achieve an average net motion along the gradient of chemoattractant by performing a biased random walk, sometimes in the form of a run-and-tumble motion, for which it is sufficient to detect the local chemoattractant concentration rather than the actual direction of its gradient. Both such tactics to perform chemotaxis (or anti-chemotactixis in case of chemorepellent) can lead to the destabilization of the suspension and produce non-trivial collective dynamics among synthetic as well as biological swimmers \cite{budrene1991complex,Theurkauff2012,Ginot2018,Lushi2012,Liebchen2017}. 
As for their living analogues \cite{Subramanian2009,Dombrowski2004} 
the dynamics of phoretic suspensions can transition from seemingly random motions to more complex collective behavior beyond a critical value of the particles' volume fraction~\cite{Theurkauff2012,Ginot2018,palacci2013living}.

Recent research efforts have focused increasingly on the control of the activity of phoretic systems at the level of their individual constituents, in order to create suspensions of active particles with individual tunable swimming speeds~\cite{Bauerle2018} or to control the formation of self-powered microgears from active particles~\cite{Aubret2018}. The individual behaviour of phoretic particles and their collective organization are related to their self-propulsion characteristics (e.g. velocity) and its robustness to external forcing. They are also critically influenced by the chemical and hydrodynamic footprints they introduce and forcings they exert on their environment,  which allow them to develop collective dynamics. These forcings, in turn, are directly and fundamentally determined by the particles' detailed shape~\cite{popescu2010,michelin2017}, size~\cite{ebbens2012,Izri2014} and surface activity and mobility distributions~\cite{Golestanian2007,Michelin2014,Lauga2016}. 
Understanding the intimate coupling of such individual design and collective behavior is the main focus of the present work and, in this paper, we aim to characterize how specific microscopic properties of individual Janus particles determine the large-scale dynamics of phoretic suspensions.

To this end, we model suspensions of chemically-active Janus particles (JPs) using a kinetic model, extending to that purpose a recent and popular framework initially proposed to study the emergence of hydrodynamic instabilities within bacterial suspension~\cite{Saintillan2008} and later extended to analyse the detailed rheology of active suspensions~\cite{Saintillan2018} or chemotaxis~ \cite{Lushi2018,Lushi2012}. The fundamental idea of such modelling is to describe the evolution in time of a probability density to find a particle  at a given location and with a particular orientation in terms of ambient mean chemical and hydrodynamic fields, which are in turn forced by the distribution and action of the particles. This approach is fundamentally limited to dilute suspensions as the coupling of two particles' dynamics is only accounted for through their action on the mediating flow and chemical fields, although corrections to account for steric interactions can also be introduced~\cite{Saintillan2013}. This generic approach is adapted here to account for the detailed individual properties of Janus particles, and how they influence individually the different fundamental parameters of the system characterizing their behavior (e.g. swimming speed, chemotactic or anti-chemotactic character). 

The rest of the paper is organized as follows. Section~\ref{Sec_Model} introduces the modeling approach chosen here, obtaining the different characteristics of individual particles and how they react to external chemical and hydrodynamic fields directly in terms of their surface properties. In a second step, these individual properties are introduced into the generic kinetic model at the suspension level, obtaining a closed set of equations driving the joint dynamics of the particle density distribution and the mean chemical and hydrodynamic fields. This model is then applied to analyze the linear stability of isotropic suspensions which are characterized initially by uniform solute and particle distributions and no hydrodynamic flow (Sec.~\ref{sec:stab_an}); for spherical particles, chemical coupling and resulting instabilities are dominant in that linear regime.  
The role of hydrodynamic interactions in setting the characteristics of the nonlinear regime arising from these instabilities, as well as their interplay with chemical interactions, are investigated using numerical simulations of the full kinetic model (Sec.~\ref{sec:num_sim}). Section~\ref{sec:conclusions} finally summarizes the main conclusions of this analysis and presents further perspectives.
 %
 %

%
%
\section{Model} \label{Sec_Model}
The goal of the present Section is to obtain a \emph{minimal} kinetic model for a dilute suspension from the detailed understanding and modeling of the chemical and hydrodynamic fields around a single phoretic Janus particle, as well as the particle self-propulsion dynamics and its response to outer chemical and hydrodynamic forcings. In Section~\ref{sec:part_level}, the individual particle properties are obtained in terms of physico-chemical characteristics before being included in the suspension model in Section~\ref{sec:susp_level}.
\subsection{Single-particle dynamics} \label{sec:part_level}
We first analyze the motion and chemo-hydrodynamic signatures (i.e. the generated chemical and hydrodynamic fields) of a single Janus particle (JP), which drifts and rotates due to the presence of a non-uniform physico-chemical field, denoted by $C$. This field can result from either the polar chemical activity of the colloid (self-propulsion) or can be externally imposed, for example by an external chemical field or by the presence of other active particles nearby (passive phoretic drift). 

A spherical half-coated Janus particle of radius $R$ is considered; at the particle scale, i.e. focusing on the concentration field in the particle's vicinity, $r\gtrsim R$, the solute dynamics is purely diffusive, i.e. its excess concentration with respect to the chemical equilibrium far from the particles satisfies Laplace's equation,
\begin{eqnarray}
   D_c\nabla^2_x C = 0, \label{Poisson_C}
\end{eqnarray}
where $D_c$ is the solute diffusion coefficient and $\nabla_x$ denotes the spatial gradient. Convective or intrinsic restoring dynamics to the background chemical equilibrium $C_\infty$, which are neglected here at the particle scale, may however be significant at the scale of the suspension, i.e. for $r\gg R$, as discussed further in Section~\ref{subsec:non_dim}. 
Local solute gradients along the particle's surface generate a local flow forcing within a thin boundary layer surrounding the particle, because of the difference in physico-chemical affinity of the solute and solvent molecules with the particle's surface~\cite{anderson89}. This phoretic effect is responsible for an apparent slip velocity on the surface of the particle \cite{Golestanian2007,Michelin2014}
\begin{eqnarray}
   \textbf{u}^* = M(\textbf{n})(\textbf{I}-\textbf{nn})\cdot  \nabla C|_{r = R}, \label{u_slip}
\end{eqnarray}
which couples the chemical and hydrodynamic fields through the mobility coefficient $M(\textbf{n})$, a physico-chemical property of the surface of the colloid which determines the particle's repulsive or attractive interaction to the solute molecules~\cite{anderson89}. Here, $\mathbf{n}=\mathbf{r}/r$ is the normal unit vector at the particle's surface $r=R$. We note that Eq.~\eqref{u_slip} is rigourosly valid only for self-diffusiophoresis in uncharged neutral electrolytes, while self-diffusiophoresis in charged electrolytes (or self-thermophoresis) would result instead in a slip velocity proportional to $\nabla \ln C|_{r = R}$~\cite{anderson89}. This logarithmic dependence may have a significant influence on the quantitative predictions of the particles' velocity when the magnitude of activity-induced concentration fluctuations are of the same order as or larger than the absolute concentration level~\cite{yang2019}. However, Eq.~\eqref{u_slip} may still provide a useful approximation in the case of charged solute molecules when fluctuations in the concentration are small compared to the absolute solute concentration level and is rigorously valid for linear stability analysis around a uniform steady state, i.e. when $C=C_0+\delta C$ with $|\delta C|\ll C_0$ (Section~\ref{sec:stab_an}).

Considering an isolated particle in unbounded flow, the reciprocal theorem for Stokes’ flow can be used to obtain from $\mathbf{u}^*$ the translational and rotational velocities $\mathbf{U}$ and $\boldsymbol\Omega$ of the colloid~\cite{Stone1996} as well as its dominant far field hydrodynamic signature, i.e. a stresslet $\mathbf{S}$ or symmetric force-dipole (the particle is force- and torque-free)~\cite{Lauga2016}, as
\begin{eqnarray}
\mathbf{U}=&-\langle\mathbf{u}^*\rangle,\qquad\boldsymbol\Omega=-\frac{3}{2R}\langle\mathbf{n}\times\mathbf{u}^*\rangle,\label{Ind_vel}\\ \mathbf{S}&=-\frac{5\eta}{2}\int_{\partial S}(\textbf{n}\textbf{u}^* + \textbf{u}^*\textbf{n})\mathrm{d}A. \label{Ind_stresslet}
\end{eqnarray}
This approach is used successively to determine the self-propulsion of a chemically-active swimmer in response to its own chemical activity in Section~\ref{self_q} and its drift dynamics in an externally-imposed chemical field in Section~\ref{ind_q}.  Exploiting the linearity of the diffusion and hydrodynamic problems, the complete dynamics (i.e. self-propulsion in an externally-imposed chemical field) is obtained by superposition of the two sets of results.
%
\subsubsection{Self-propulsion velocity and self-induced stresslet} \label{self_q}
The response of the Janus particle to its own chemical activity is obtained by solving Eq.~\eqref{Poisson_C} together with the boundary conditions
\begin{eqnarray}
   D_c  \textbf{n} \cdot \nabla C(r)|_{r = R} = -A(\textbf{n}) \ \ \textrm{and} \ \ C|_{r\rightarrow{}\infty} \rightarrow{} 0, \label{Poisson_C_BC}
\end{eqnarray}
where $A(\textbf{n})$ is the rate of production of the chemical solute at the colloid's boundary and quantifies its chemical activity. For simplicity, we assume here that it takes the form of a fixed-flux solute release ($A>0$) or consumption ($A<0$), but a more general (and concentration-dependent) form of $A$ could be considered to account for more complex surface kinetics~\cite{ebbens2012,Saha2014,Tatulea-Codrean2018}. In the following, we focus on hemispheric Janus particles with piecewise uniform activity $A(\textbf{n})=A_b$ on their back side ($\mathbf{n}\cdot\mathbf{p}<0$) and $A(\textbf{n})=A_f$ on their front side ($\mathbf{n}\cdot\mathbf{p}>0$) with $\mathbf{p}$ the unit vector pointing toward the front of the particle and along its axis of symmetry. We further define $A^+=A_b+A_f$ and $A^- = A_b-A_f$, respectively the total activity and activity contrast, and adopt the same definitions for the mobility equivalents, $M^+$ and $M^-$. In the following, we also assume that the particles act as net sources, so that $A^+>0$. The solution to the Laplace problem in equations~\eqref{Poisson_C} and~\eqref{Poisson_C_BC} is obtained as a series of spherical harmonics \cite{Kanso2019,Golestanian2007,VarmaMichelin_SoftMatter,Varma2019} 
\begin{eqnarray}
  C = \left( \frac{2\pi R^2 A^+}{D_c} \right)\frac{1}{4\pi r} - \left( \frac{3\pi R^3 A^-}{2 D_c} \right) \left( \frac{\mathbf{p}\cdot\mathbf{r}}{4\pi r^3} \right) + \sum_{m=2}^{\infty} \frac{A_mR}{(m+1)D_c} \left(\frac{R}{r}\right)^{m+1} P_m(\mu) ,  \label{C_exp} 
\end{eqnarray}
with $\mu=\textbf{p}\cdot\textbf{r}/r $ and $P_m$ the $m^\textrm{th}$ Legendre polynomial. The coefficients $A_m$ are the Legendre projections of the activity distribution, i.e. $A_m= \frac{2m+1}{2} \int_{-1}^{1} A(\mu)P_m(\mu)\textrm{d}\mu $. The slowest decaying terms in this expansion provide the chemical signature of the particle and include (i) a source of solute proportional to the net
production rate $A^+$ and (ii) a source dipole proportional
to $A^-$. We note that, for hemispheric swimmers, $A_m=0$ for even $m$.

From Eq.~\eqref{u_slip}, the resulting nonuniform distribution of solute at the surface of the colloid generates a slip velocity 
\begin{equation}
\mathbf{u}^*_s=M(\mathbf{n})\left[ - \frac{3 A^-}{8 D_c}+ \sum_{m=2}^{\infty} \frac{A_m}{(m+1)D_c} P_m'(\mu)\right]\left(\mathbf{I}-\mathbf{n}\mathbf{n}\right)\cdot \mathbf{p}. \label{u_slip_2}
\end{equation}
Substitution into Eq.~\eqref{Ind_vel} provides the self-propulsion velocity $\mathbf{U}_s=U_0\mathbf{p}$, with 
\begin{eqnarray}
  U_0 = \frac{A^-M^+}{8D_c}, \label{SelfP_dim}
\end{eqnarray}
and the self-rotation velocity vanishes due to the problem's symmetry, $\boldsymbol\Omega_s=0$.%

The stresslet associated with the particle self-propulsion is then obtained from Eqs.~\eqref{Ind_stresslet} and \eqref{u_slip} using the problem's axisymmetry, as 
\begin{equation}
\boldsymbol\sigma_s=\sigma_s\left(\mathbf{p}\mathbf{p}-\frac{\mathbf{I}}{3}\right),\qquad \textrm{with   }
        \sigma_s = - \frac{10\pi \eta a^2 \kappa M^- A^-}{D_c} \hspace{3mm}
        \label{sigma_s}
  \end{equation}
where $\kappa$ is a numerical constant obtained as
\begin{eqnarray}
   \kappa = \frac{3}{4}\sum_{m=1}^{\infty} \frac{2m+1}{m+1} \left[ \int_0^1P_m \textrm{d}\mu \right] \left[ \int_0^1 \mu (1-\mu^2)P'_m \textrm{d}\mu \right]\approx 0.0872.
\end{eqnarray}
%
\subsubsection{Externally-induced drift, rotation and stresslet} \label{ind_q}
Because their mobility property gives them the ability to generate slip in response to any surface concentration gradient, regardless of its origin, phoretic particles may also drift in externally-imposed non-homogeneous concentration fields, in particular those resulting from the presence of other active particles. 
To obtain the resulting translational and rotational drifts as well as the particle's hydrodynamic signature, it is equivalent to consider the problem of a chemically-passive particle (i.e. $A(\textbf{n})=0$) immersed in a non-uniform concentration field. In the following, we will focus on dilute suspensions where the concentration signature induced by other particles will vary slowly around each particle. As a result, we focus here on a slowly-varying externally-imposed field in the absence of the particle, i.e. $C_\textrm{ext}(\mathbf{x})\sim C_\infty+\mathbf{G}\cdot\mathbf{x}$, where the concentration gradient $\mathbf{G}=\nabla C_\textrm{ext}$ is considered uniform and equal to its value at the particle's centroid. Note that by doing so we neglect any second order gradient and quadratic variations of the external field which is reasonable for dilute suspensions as such correction would be $O(R/l_c)$ or smaller, with $l_c$ the characteristic length scale describing the suspension dynamics (see Sec.~\ref{sec:susp_level}).

The presence of the particle modifies this concentration distribution, which now writes for $|\textbf{r}|\geq R$ as
   \begin{eqnarray}
        C = C_{\infty} + \textbf{G}\cdot\textbf{r} \left( 1+\frac{R^3}{2r^3} \right)
   \end{eqnarray}
and the resulting induced slip velocity is
   \begin{eqnarray}
      \textbf{u}^*_{i} = \frac{3}{2}M(\textbf{n})(\textbf{I}-\textbf{n}\textbf{n})\cdot \textbf{G}.  \label{u_s_ind}
   \end{eqnarray}

In response to this slip velocity $\textbf{u}^*_{i}$, the hemispheric Janus particle's induced translational and rotational drifts are $\mathbf{U}_i=\chi_t\textbf{G}$ and $\boldsymbol\Omega_i=\chi_r \textbf{p} \times \textbf{G}$, where $\chi_t$ and $\chi_r$ read
   \begin{eqnarray}
          \chi_t = -\frac{M^+}{2}, \qquad \chi_r = \frac{9}{16} \frac{M^-}{R}\cdot
   \end{eqnarray}
Note that no translational motion parallel to $\textbf{p}$ is induced for a half-coated (i.e. hemispheric) colloid in contrast with more generic particles~\cite{Tatulea-Codrean2018,Kanso2019}.

To obtain the induced stresslet $\boldsymbol\sigma_i$, Eq.~\eqref{u_s_ind} is substituted into Eq.~\eqref{Ind_stresslet}, yielding
\begin{eqnarray}
    \boldsymbol{\sigma}_i = -\frac{15\eta}{4}  \int_{\partial S} M(\textbf{n}) \Big\{ \left[ (\textbf{I}-\textbf{nn}) \cdot \textbf{G} \right] \textbf{n} + \textbf{n} \left[ (\textbf{I}-\textbf{nn}) \cdot \textbf{G} \right] \Big\} \textrm{d}S \label{deriv_Sind_1}.
\end{eqnarray}
For the case of hemispherical swimmers the mobility is $M(\textbf{n})=M_f$ for $\textbf{n}\cdot\textbf{p}>0$ and $M(\textbf{n})=M_b$ for $\textbf{n}\cdot\textbf{p}<0$; in that case, the integral in Eq.~\eqref{deriv_Sind_1} can be conveniently rewritten as the sum of two contribution on each hemisphere using the following results
\begin{align}
\int_{\mathbf{n}\cdot\mathbf{p}>0}\mathbf{n}\,\mathrm{d}S=\pi R^2\mathbf{p},\qquad \int_{\mathbf{n}\cdot\mathbf{p}>0}\mathbf{nnn}\,\mathrm{d}S=\frac{\pi R^2\mathbf{p}}{4}\left[\mathbf{p}\mathbf{I}+\mathbf{I}\mathbf{p}+(\mathbf{I}\mathbf{p})^{T_{23}}\right]
\end{align}
where $\mathbf{A}^{T_{23}}$ is the transpose of the third-order tensor $\mathbf{A}$ with respect to its last two indices. The integrals on $\mathbf{n}\cdot\mathbf{p}<0$ are obtained by changing $\mathbf{p}$ into $-\mathbf{p}$, and the induced stresslet is finally computed as 
   \begin{equation}
        \boldsymbol{\sigma}_i  = \sigma_i \left[ \textbf{G}\textbf{p} + \textbf{p}\textbf{G} + (\textbf{G}\cdot\textbf{p}) (\textbf{pp}-\textbf{I}) \right],\qquad \textrm{with    } \sigma_i = \frac{15}{8}\eta R^2 \pi M^-.\label{Ind_stresslet2}
   \end{equation}
   
%
%
\subsubsection{Individual particle properties: a note on the role of the physico-chemical property} \label{individual_char}
The self-propulsion and induced-drift velocities and stresslets were obtained in Sections~\ref{self_q} and \ref{ind_q} from the detailed chemical dynamics at the particle level explicitly in terms of the particle's activity and mobility properties (i.e. $M^+$, $M^-$, $A^+$ and $A^-$) in the case of hemispheric particles.

Specifically, it should be noted that the self-propulsion velocity and self-generated stresslet are both proportional to the front-back activity contrast, $A^-$, as it is responsible for the self-generated chemical gradient at the surface of an isolated particle.  
In contrast, externally-induced drifts and stresslet are proportional to the magnitude of the externally-imposed gradient. In a suspension, where the external concentration field results from the dominant chemical signature of other particles (i.e. a net source of intensity $A^+$, see Eq.~\eqref{C_exp}), these quantities are therefore proportional to the \emph{mean} activity, $A^+$.

Each velocity or stresslet intensity is also a linear function of the mobility distribution. Namely, the translational velocities, either self-induced ($u_0$) or externally-induced ($\sim \chi_t$), are proportional to the \emph{mean} mobility, $M^+$, as the latter determines the average slip velocity on the surface of the colloid. In contrast, the (self- and externally-induced) stresslets of the particle, $\sigma_i$ and  $\sigma_s$, as well as its rotational drift velocity, $\chi_r$, are proportional to the front-back mobility \emph{contrast}, $M^-$.
%

%
%
%
%
%
%
\subsection{Kinetic model for suspension dynamics}
\label{sec:susp_level}
\subsubsection{Governing equations}
Having understood and fully-characterized the behavior and chemo-hydrodynamic footprints of individual particles, we now turn to the description of a dilute suspension of auto-phoretic Janus swimmers. The approach followed here considers that the suspension dynamics are studied on a length scale much larger than the particle radius, and instead of characterizing each particle's state individually the probability to find a particle in a given small volume of fluid with a set orientation is fully described by the probability distribution function $\Psi(\textbf{x},\textbf{p},t)$ of
the particle position, $\textbf{x}$, and director, $\textbf{p}$ \cite{Saintillan2008,Lushi2018,Lushi2012}. The evolution
of the suspension then classically follows a Smoluchowski equation    
\begin{eqnarray}
    \frac{\partial \Psi}{\partial t} &= -\nabla_x\cdot(\Psi \dot{\textbf{x}}) - \nabla_p\cdot(\Psi \dot{\textbf{p}}) , \label{EvolEqPsi_dim} 
\end{eqnarray}
where $\nabla_p$ denotes the gradient operator on the unit sphere. The distribution function is normalized so that \cite{Saintillan2008}
\begin{eqnarray}
   \frac{1}{\textrm{V}} \int_{\textrm{V}}\textrm{d}\textbf{x}\int_{\textrm{S}}\textrm{d}\textbf{p} \Psi(\textbf{x},\textbf{p},t) = n ,
\end{eqnarray}
where $n=N/V$ is the mean number density of particle in the suspension and $N$ is the total number of particles within the volume of interest, $V=L^3$. 

The translational and rotational fluxes, $\dot{\textbf{x}}$ and $ \dot{\textbf{p}}$, are obtained from the corresponding deterministic velocities of an individual particle located at $\mathbf{x}$ and oriented along $\mathbf{p}$ in response to its own activity and to the hydrodynamic and phoretic mean fields, $\textbf{u}(\textbf{x},t)$ and $C(\textbf{x},t)$ in its vicinity, respectively. In this dilute limit, these fluxes are directly expressed by superimposing the self-propulsion and induced drifts of an individual JP determined in Section~\ref{sec:part_level} as well as the leading-order classical Faxen's law for a spherical particle, yielding
\begin{eqnarray}
    \dot{\textbf{x}} &=& U_0\textbf{p} + \textbf{u} + \chi_t\nabla_x C - D_x\nabla_x(\ln(\Psi)) ,  \label{xdot_dim} \\
    \dot{\textbf{p}} &=& \frac{1}{2}\boldsymbol{\omega}\times\textbf{p}  + \chi_r( \textbf{p} \times \nabla_x C )\times \textbf{p} - D_p\nabla_p(\ln(\Psi)) ,  \label{pdot_dim}
\end{eqnarray}
where $\boldsymbol{\omega} = \nabla_x\times\textbf{u}$ is the vorticity vector. The last terms in Eqs.~\eqref{xdot_dim} and \eqref{pdot_dim} account for the translational and rotational diffusion of the particles with constant diffusion coefficients $D_x$ and $D_p$, respectively, and model the integral effect of the thermal noise of the bath in the over-damped regime. 
Due to the fully deterministic modeling of the chemically-induced rotation which is well-suited for phoretic particles, this approach is sometimes referred to as turning-particle model. For other systems such as swimming bacteria, other models have been proposed (e.g. run-and-tumble~\cite{Lushi2018}).   

The mean pressure and velocity fields in the suspension, $q$ and $\mathbf{u}$, satisfy the incompressible Stokes equations, forced by the hydrodynamic stresses generated by each JP individually,
\begin{eqnarray}
    \nabla_x\cdot\textbf{u} &=& 0 ,  \label{continuity_dim} \\ 
    -\eta\nabla^2_x\textbf{u} + \nabla_x q &=& \nabla_x\cdot \boldsymbol{\Sigma}.   \label{stokeseq_dim}
\end{eqnarray}
In the mean field description of a dilute suspension, the bulk effect of the swimmers is described by superimposing the active stresses produced by different swimmers at a given location , \eqref{sigma_s} and \eqref{Ind_stresslet2}, i.e.
  \begin{align}
      \boldsymbol{\Sigma}(\textbf{x},t) = \int_S  \boldsymbol{\sigma}_s \Psi(\textbf{x},\textbf{p},t) \textrm{d}\textbf{p} + \int_S  \boldsymbol{\sigma}_i \Psi(\textbf{x},\textbf{p},t) \textrm{d}\textbf{p} .
  \end{align}

At the suspension scale, the solute produced by each swimmer (at a rate $2 \pi R^2A^+$) diffuses and may also be advected by the fluid flow. We further account physically for the finite-time intrinsic relaxation rate ($\beta_1$) of the chemical system toward its background equilibrium far from all active particles.
As a result, the equation governing the dynamics of the solute concentration, $C$,  reads
\begin{align}
    \frac{\partial C}{\partial t} + \textbf{u}\cdot\nabla_x C = D_c\nabla_x^2 C - \beta_1 C + 2\pi R^2 A^+ \Phi,      \label{conceq_dim}
\end{align}
where $\Phi(\textbf{x},t) = \int_S \Psi(\textbf{x},\textbf{p},t)\textrm{d}\textbf{p}$ is the particle density. The last term on the RHS of Eq.~\eqref{conceq_dim} is a coarsed-grained representation of the production of $C$ due to the presence of swimmers, which are here considered to be net sources ($A^+>0$).

\subsubsection{Nondimensional equations} \label{subsec:non_dim}
The governing equations are made dimensionless using the reference length scale $l_c=(nR^2)^{-1}$ introduced by \cite{Saintillan2008} for such suspensions. Note that with this choice of $l_c$, the nondimensional particle radius $\varphi = R/l_c = 3\nu/(4\pi)$ is proportional to the volume fraction occupied by the swimmers, $\nu=NV_p/V$ with $V_p=\frac{4}{3}\pi R^3$. The associated time scale, $t_c = l_c^2/D_c$, is based on the diffusion time of the solute. Finally, the characteristic concentration scale  is obtained as $C_c = l_c A^+ / D_c$ by the balance of chemical production by the phoretic particles ($nR^2A^+$) and the diffusive flux at the suspension level ($D_cC_c/l_c^2$).

The nondimensional fluxes become
\begin{eqnarray}
     \dot{\textbf{x}} &=& u_0\textbf{p} + \textbf{u} + \xi_t\nabla_x C - d_x\nabla_x(\ln(\Psi)) ,  \label{xdot} \\
    \dot{\textbf{p}} &=& \frac{1}{2}\boldsymbol{\omega}\times\textbf{p}  + \frac{\xi_r}{\varphi}( \textbf{p} \times \nabla_x C )\times \textbf{p} - d_p\nabla_p(\ln(\Psi)) ,  \label{pdot}
\end{eqnarray}
where the non-dimensional self-propulsion and chemically-induced drifts are obtained from the dimensional properties of the particles as 
\begin{eqnarray}
   u_0 = \frac{A^{-}M^{+}}{8D^{2}_c n R^2 } , \qquad
   \xi_t = -\frac{M^+ A^+}{n R^2 D_c^2}, \qquad
   \xi_r = \frac{9}{16}\frac{M^- A^+}{D_c^2 n R^2}\cdot
\end{eqnarray}
The reduced diffusion coefficients are respectively defined as $d_x=D_x/D_c$ and $d_p=D_p l_c^2/D_c$.

We will treat here $\xi_t$ and $\xi_r$ as independent non-dimensional measures of the mean mobility  $M^+$ and mobility contrast $M^-$, respectively. The self-propulsion velocity in turn, can be seen as the non-dimensional measure of the activity contrast $A^-$, although it is also proportional to $\xi_t$. Consequently, it is not physically relevant to consider $\xi_t=0$ and $u_0 \neq 0$, while the reverse situation ($u_0 =0$ and $\xi_t\neq 0$) corresponds to a particle of uniform activity. 

 It should therefore be noted that the nondimensional stresslet intensities are not independent parameters but instead can be expressed in terms of the others as
\begin{align}
    \alpha_s = 10\pi\kappa\frac{256}{9}\frac{u_0 \xi_r}{\xi_t} \hspace{0.3cm} \textrm{and} \hspace{0.3cm}
    \alpha_i =  \frac{30}{9}\pi \xi_r , \label{nondim_stress_int}
\end{align} showing the link between the velocities induced by chemical interactions and the strength of the hydrodynamic forcing exerted by the swimmers.
Noticeably, the coefficient for the rotational velocity induced by the phoretic field, $\xi_r/\varphi$, is the only parameter that depends on the volume fraction and is a consequence of the different scaling of the two hydrodynamic and chemical rotational drift with the inter-particle distance.  

The conservation equation~\eqref{EvolEqPsi_dim} remains unchanged with the distribution function normalized by the mean particle density
    \begin{eqnarray}
       \frac{1}{\textrm{V}} \int_{\textrm{V}}\textrm{d}\textbf{x}\int_{\textrm{S}}\textrm{d}\textbf{p} \Psi(\textbf{x},\textbf{p},t) = 1 ,
    \end{eqnarray}
where $V=(L/l_c)^3$ and $\Psi$ has conserved mean $1/4\pi$.
     
The continuity and momentum equations become
    \begin{eqnarray}
    \nabla_x\cdot\textbf{u} &=& 0 ,\label{conteq} \\
    -\nabla^2_x\textbf{u} + \nabla_x q &=& \nabla_x\cdot \boldsymbol{\Sigma} . \label{momeq}
\end{eqnarray}

Finally, the concentration equation becomes  in non-dimensional form 
\begin{align}
   \frac{\partial C}{\partial t} &+ \textbf{u}\cdot\nabla_xC = 
    \nabla_x^2C - \beta C + 2\pi \Phi , \label{conceq} 
\end{align}
where $\beta^{-1/2}=l^*/l_c$ is the reduced screening length $l^*=\sqrt{D_c/\beta_1}$ introduced at the scale of the suspension by the finite-time intrinsic relaxation of the chemical system toward its background equilibrium.

In the following, we assume that $\beta=O(1)$ which guarantees the existence of a steady state solution for an isotropic suspension and implies that the relaxation toward chemical equilibrium away from active particles occurs at a finite distance that is much larger than the particle size (thus allowing for chemical interactions between particles).  This effective relaxation $-\beta C$, introduces an exponential decay of the concentration field away from chemical sources (rather than the algebraic one associated to diffusion) with a $l^*=O(l_c)$ characteristic screening length.  As a result such exponential screening is negligible at the particle scale (i.e. at a $O(R)$ distance from the source particle) and the degradation term could indeed be neglected in the derivations of Section~\ref{sec:part_level}. Similarly, the convection of solute by the fluid flow (i.e. left-hand-side of Eq.~\eqref{conceq}) plays an $O(1)$ role at the suspension scale $l_c$ but is negligible at the scale of the particle radius, where the solute dynamics thus simply satisfies Laplace's equation.

%
%
%
%
\section{Linear stability analysis of a nearly isotropic suspension} \label{sec:stab_an}
The previous system admits a trivial uniform equilibrium solution where the distribution of particles is homogeneous and isotropic, i.e. $\Psi(\mathbf{x},\mathbf{p},t)=\Psi_0=1/4\pi$, $\Phi(\mathbf{x},t)=\Phi_0=1$ and $C(\mathbf{x},t)=C_0$ with  $C_0=2\pi\Phi_0/\beta$. In this section, we analyze the stability of linear perturbations of this isotropic solution, i.e. by expanding $\Psi(\textbf{x},\textbf{p},t) = \frac{1}{4\pi} \left[ 1+\delta\Psi(\textbf{x},\textbf{p},t) \right] $ and $C(\textbf{x},t) = C_0 + \delta C(\mathbf{x},t)$, where $\delta f$ indicates a small perturbation of a quantity $f$.
\subsection{Dispersion relation}
 Using this expansion, the linearized governing equations read
\begin{eqnarray}
    \frac{\partial \delta\Psi}{\partial t} &=& - u_0\textbf{p}\cdot\nabla_x\delta\Psi - \xi_t\nabla_x^2\delta C + d_x\nabla_x^2\delta\Psi + 2\frac{\xi_r}{\varphi}\textbf{p}\cdot\nabla \delta C, \\
    \frac{\partial \delta C}{\partial t} &=& -\beta \delta C + 2\pi\delta\Phi + \nabla_x^2 \delta C. \label{eqClin}
\end{eqnarray}
where use is made of the identity $\nabla_p \cdot \left[ (\textbf{p} \times \nabla_x C) \times \textbf{p} \right] = -2 \textbf{p}\cdot\nabla_xC$.
Upon linearization, we seek solutions written as planar waves with wave vector $\textbf{k}$ and growth rate $\sigma$, i.e. $\delta\Psi(\textbf{x},\textbf{p},t) = \tilde{\Psi}(\textbf{k},\textbf{p})\exp(i\textbf{k}\cdot\textbf{x} + \sigma t)$ (similar definitions are used for $\tilde{C}$ and $\tilde\Phi$), yielding
\begin{eqnarray}
   ( \sigma + i  u_0\textbf{p} \cdot \textbf{k} + d_x k^2 ) \tilde{\Psi}  = 
     \left( \xi_t k^2 + 2 i \frac{\xi_r}{\varphi} \textbf{p} \cdot \textbf{k} \right)\tilde{C} \qquad \textrm{and}\qquad  \tilde{C} = \frac{2\pi}{\sigma + \beta + k^2}\tilde{\Phi}.\label{mom_cont_evolPsi_together}
\end{eqnarray}
Integration of Eq.~\eqref{mom_cont_evolPsi_together} over all orientations $\mathbf{p}$ yields
\begin{eqnarray}
    i  u_0 \tilde{\textbf{n}}\cdot\textbf{k} + 4\pi(\sigma+d_x k^2)\tilde{\Phi}  =
  \frac{8\pi^2 \xi_tk^2}{\sigma + \beta + k^2}\tilde{\Phi},  \label{1stEq}
\end{eqnarray}
where $4\pi\tilde{\Phi}$ and $\tilde{\textbf{n}}$ are respectively the particle concentration and local polarization, and are mathematically obtained as the zero-th and first moment of $\tilde\Psi$ in the orientation space. 
{We remark that, in the limit $u_0=0$, Eq.~\eqref{1stEq} relates the growth rate and wave number of fluctuations of the particle density $\Phi$, which evolve independently of the full distribution function $\Psi$ in the absence of self-propulsion.} 

Equation~\eqref{mom_cont_evolPsi_together} is then rewritten as 
\begin{eqnarray}
  \tilde{\Psi} =  \left[\frac{\xi_t k^2 + 2 i \frac{\xi_r}{\varphi} \textbf{p}\cdot\textbf{k} }{\sigma + i u_0 \textbf{p}\cdot\textbf{k}  + d_x k^2}\right]\tilde{C} \label{mom_cont_evolPsi_together_Sant}   
\end{eqnarray}
and its first moment with respect to orientation $\mathbf{p}$ can then be projected along the wave vector $\mathbf{k}$ of the considered eigenmode:
\begin{eqnarray}
     \int_S \tilde{\Psi}(\mathbf{p}\cdot\mathbf{k})\,\mathrm{d}\mathbf{p}= \Tilde{\textbf{n}} \cdot \textbf{k} = \Tilde{C} \xi_t k^2\int_S \frac{
    \textbf{p}\cdot \textbf{k} }{\sigma + i u_0 \textbf{p}\cdot \textbf{k} + d_x k^2} \textrm{d}\textbf{p} 
    +2 i  \frac{\xi_r}{\varphi} \Tilde{C} \int_S \frac{(\textbf{p}\cdot \textbf{k})^2}{{\sigma + i u_0 \textbf{p}\cdot \textbf{k} + d_x k^2}} \textrm{d}\textbf{p}. \label{Eq_intermediate_1}
\end{eqnarray}
The integrals on the RHS of Eq.~\eqref{Eq_intermediate_1} can be solved in terms of the scalar variable $\mu = \textbf{p}\cdot \textbf{k}/k$, yielding 
\begin{eqnarray}
\mathrm{i}u_0\Tilde{\textbf{n}} \cdot \textbf{k} =4\pi\left[-\xi_t k^2 +\frac{2a\xi_r k}{\varphi} \right](aI_0-1)\tilde{C}\qquad \textrm{with   }I_0(a)\equiv\frac{1}{2}\int_{-1}^1\frac{\mathrm{d}\mu}{a+\mathrm{i}\mu}=\tan^{-1}\left(\frac{1}{a}\right), \label{Eq_intermediate_2}
\end{eqnarray}
where $a \equiv (\sigma + d_x k^2)/(k u_0)$. Finally, the LHS of Eq.~\eqref{Eq_intermediate_2} is evaluated using Eq.~\eqref{1stEq}, the $\theta$-integrals are computed and $\Tilde{C}$ is expressed as a function of $\Tilde{\Phi}$ using Eq.~\eqref{mom_cont_evolPsi_together}, yielding the dispersion relation for the modes of $\Psi$ 
\begin{eqnarray}
    u_0 - \frac{2\xi_rc}{\varphi} \left( 1-a\tan^{-1}\frac{1}{a}  \right)  - \xi_t k c \tan^{-1}\frac{1}{a}= 0 , \label{DispRelChem}
\end{eqnarray}
where $c \equiv 2\pi/(\sigma+\beta +k^2)$.
Equation~\eqref{DispRelChem} generalizes to phoretic particles the dispersion relation obtained for chemical instabilities of the turning particle model applied to chemotaxis in Ref.~\cite{Lushi2018}, with the addition of the phoretic drift experienced by Janus particles in external chemical fields. 
The hydrodynamic instability, which stems from the shear-induced reorientation of elongated swimmers in the flows they generate by their forcing on the fluid~\cite{Saintillan2008}, is not observed here for spherical Janus swimmers.    

At the linear stability level, it should be noted that the coupling mechanisms present in Eq.~\eqref{DispRelChem} are solely the phoretic drift and chemical reorientation, i.e. are only linked to the chemical interactions of the particles. Hydrodynamic interactions are not present due to the spherical shape of the particles and they are thus only expected to play a role in a later phase of saturation of any potential instability. This influence will be discussed in Section~\ref{sec:num_sim}.

In the following, we analyze three possible routes to instability of an isotropic suspension corresponding to three types of combinations of the particles' chemical properties. These are (i) a positive phoretic attraction ($\xi_t>0$) with no chemical reorientation ($\xi_r\approx 0$) or \emph{phoretic limit} (Section~\ref{phoretic_inst}), (ii) a positive chemical reorientation ($\xi_r>0$) or \emph{chemotactic limit} (Section~\ref{chemo_inst}) and (iii) a negative chemical reorientation ($\xi_r<0$) or \emph{anti-chemotactic limit} (Section~\ref{delay_inst}). The last two regimes are investigated for negative phoretic drift ($\xi_t<0$) to isolate the effect of chemical reorientation from that of phoretic clustering investigated in case (i).

For completeness, we finally remark that another class of instability of phoretic particles, the so-called \emph{Janus instability} was recently reported~\cite{Liebchen2015,Liebchen2017} using a coarse-grained model that does not include any hydrodynamic coupling of the particles but accounts for the first two leading order terms in their chemical signatures, namely a source and dipole, the latter describing their chemical polarity. Such chemical polarity is in fact a key ingredient to the Janus instability, which is therefore not observed in the present framework where chemical and hydrodynamic interactions are both represented to a consistent level of asymptotic approximation: hydrodynamic interactions and chemical interactions through the source signature indeed induce $1/d^2$ particle velocities while the chemical polarity's dipolar signature leads to $1/d^3$ velocities with $d$ the interparticle distance. As a result, such dipolar interactions are not dominant in the far-field framework of the dilute regime but may become of significant importance at higher volume fractions which are beyond the scope of the present work.

%
%
%
%
\subsection{Phoretic limit} \label{phoretic_inst}
\begin{figure}[!htb]
  \includegraphics[scale= 0.4]{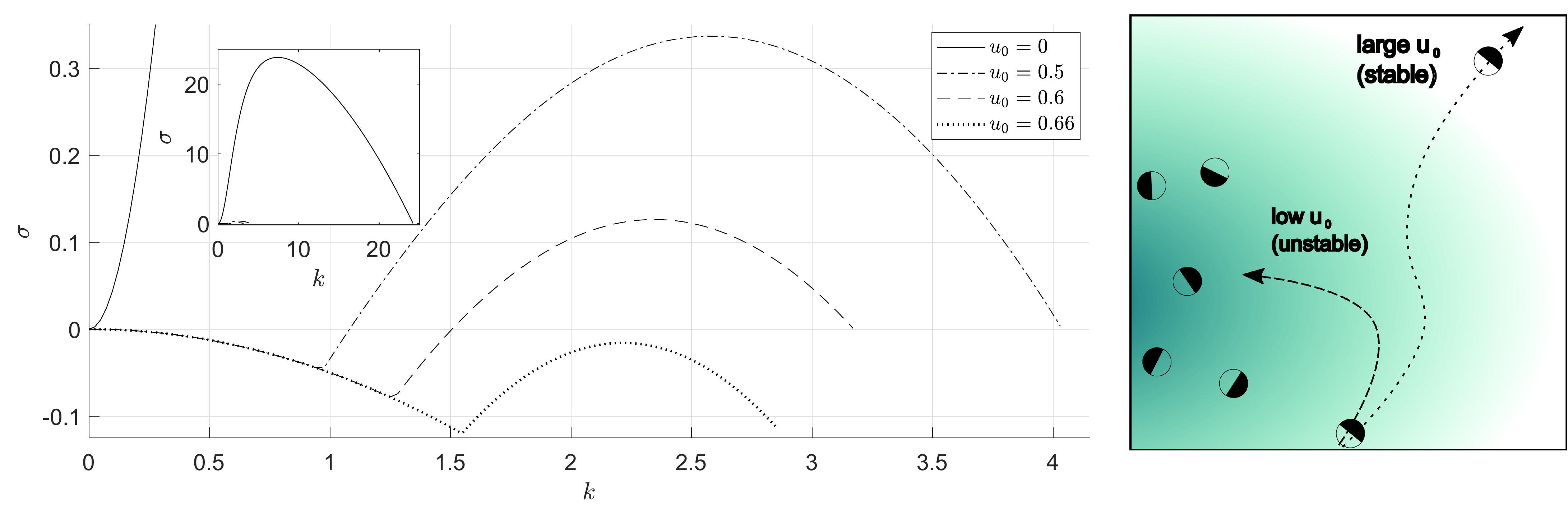}
  \caption{Phoretic limit. Left: Evolution of the growth rate of the least stable mode as obtained numerically from Eq.\eqref{DispRelChem} in the phoretic limit ($\xi_r=0$ and $\xi_t=0.375$) for different values of $u_0$ with $\beta=2\pi$ and $d_x=0.05$. The inset shows the same quantity with different axis' range to include the case where self-propulsion is absent ($u_0=0$, Eq.~\eqref{DispRelMi_noSP}). Right: Schematic representation of the competition between externally induced drift and self-propulsion in the phoretic limit.  } \label{Mi_inst}
\end{figure}
In order to analyze the phoretic limit specifically, we focus in this section on particles with (i) negligible chemical reorientation ability, i.e. particles with zero or small mobility contrast ($M^- \rightarrow 0$ so that $\xi_r\rightarrow 0$) and (ii) negative mean mobility ($M^+<0$), resulting in phoretic attraction of the chemical-emitting particles ($\xi_t > 0$):
A local accumulation of particles produces an  attractive chemical gradient that promotes the migration of other colloids towards it. As more particles reach that region the concentration of chemoattractant increases even further leading to a positive feedback and the instability of the system. 
The induced chemical drift at the base of the destabilizing mechanism competes with self-propulsion. As depicted in Fig.~\ref{Mi_inst} (right), the ability of the attractive chemical gradient to trap other particles can be reduced, if not compromised, for fast swimmers that may swim past the chemical-rich region. 

The stabilizing effect of self-propulsion in the phoretic limit can be quantified by solving the chemical dispersion relation for different values of $u_0$. 
We first focus on the simpler case where the particle does not self-propel, by setting $u_0=0$ in Eq.~\eqref{1stEq}. 
For conciseness of presentation here, and having checked that this does not alter the conclusions, we consider the limit of a quasi-steady phoretic field, namely neglecting the unsteady term $\partial C/\partial t =0$ in Eq.~\eqref{eqClin}. 
The dispersion relation can then be solved analytically, as (see inset Fig.~\ref{Mi_inst})
   \begin{eqnarray}
        \sigma = \frac{8\pi^2 \xi_t k^2}{\beta + k^2} - d_x k^2 . \label{DispRelMi_noSP}
   \end{eqnarray}
Long waves (small $k$) are unstable for all $0\leq k\leq k_c$ with $k_c= \sqrt{(-\beta d_x + 8\pi^2 \xi_t)/d_x}$ and the most unstable mode is $ k_{M} = \left( -\beta + 2\sqrt{2\pi^2\beta \xi_t /d_x } \right)^{1/2}$.
We observe that, in principle, the effect of particle diffusion can suppress the instability. An estimate of the value of $d_x=D_x/D_c$ being $O(10^{-2})$ or less can be obtained from the experimentally measured translational diffusion of Janus particles \cite{Theurkauff2012,Buttinoni2013}, suggesting that in practice the effect of particle diffusion is far too limited to stabilize the suspension. 
Particle diffusion however plays an important role in the stabilization of short waves (large $k$).  

When $u_0 \neq 0$, the dispersion relation for modes of inhomogeneous particle density can not be obtained directly from Eq.~\eqref{1stEq}. Instead, Eq.~\eqref{DispRelChem} expanded for small $k$, keeping $\xi_t=0$, leads to
   \begin{eqnarray}
   \sigma = -d_x k^2 + O(k^4), \label{Low_k_Mi}
   \end{eqnarray}
which shows that long wavelength modes are now stable.
 
In the general case, Eq.~\eqref{DispRelChem} is  solved numerically to obtain the solution for finite wavenumbers (see Fig.~\ref{Mi_inst}), confirming that self-propulsion has a stabilizing effect on the suspension.
At the heart of the destabilizing mechanism lays the strength of the attractive phoretic drift, which is proportional to the magnitude of the chemical gradient and therefore vanishes for arbitrary long wavelength (i.e. $k\rightarrow 0$). For this reason, any non-zero value of $u_0$ stabilises long-wavelength modes. By further increasing the value of $u_0$ the instability can be suppressed for any $k$ (see Fig.~\ref{Mi_inst}).

The gold-platinum colloids used in \cite{Theurkauff2012, Ginot2018} are experimental examples of phoretically-attractive particles with a mobility that is close to uniform (uniform $\zeta$-potential), so that their dynamics should correspond to the phoretic limit considered here. Particle aggregates that are small compared to the size of the suspension are observed to form and never coalesce~\citep{Theurkauff2012}, suggesting the presence of a wave-selective mechanism (preferred wave number) as the one just described. 
Moreover, the gas phase in between clusters could be seen itself as a local more dilute suspension, which we argue could be completely stabilized by the effect of self-propulsion. It is noteworthy that the present model can describe the behavior of the suspension only where it is locally dilute because of the far-field approximation. Consequently, it can not capture the mechanism responsible for the increasing size of the clusters with $u_0$, as this relies on steric interactions~\citep{Buttinoni2013}.       

A similar mechanism to the one discussed in this section was also analysed in the seminal work of Ref. \cite{Keller1970} to model chemotaxis of microorganisms.   
The model used in \cite{Keller1970}  
describes the evolution of the density of swimmers, an equivalent to $\Phi(\textbf{x},t)$, while the polarity of the suspension is not described. The kinetic model used here traces also the probability distribution in the orientational space, therefore revealing the above-mentioned stabilizing and wave-selective effects of self-propulsion.
\subsection{Chemotactic limit}  \label{chemo_inst}
\begin{figure}
        \includegraphics[scale= 0.37]{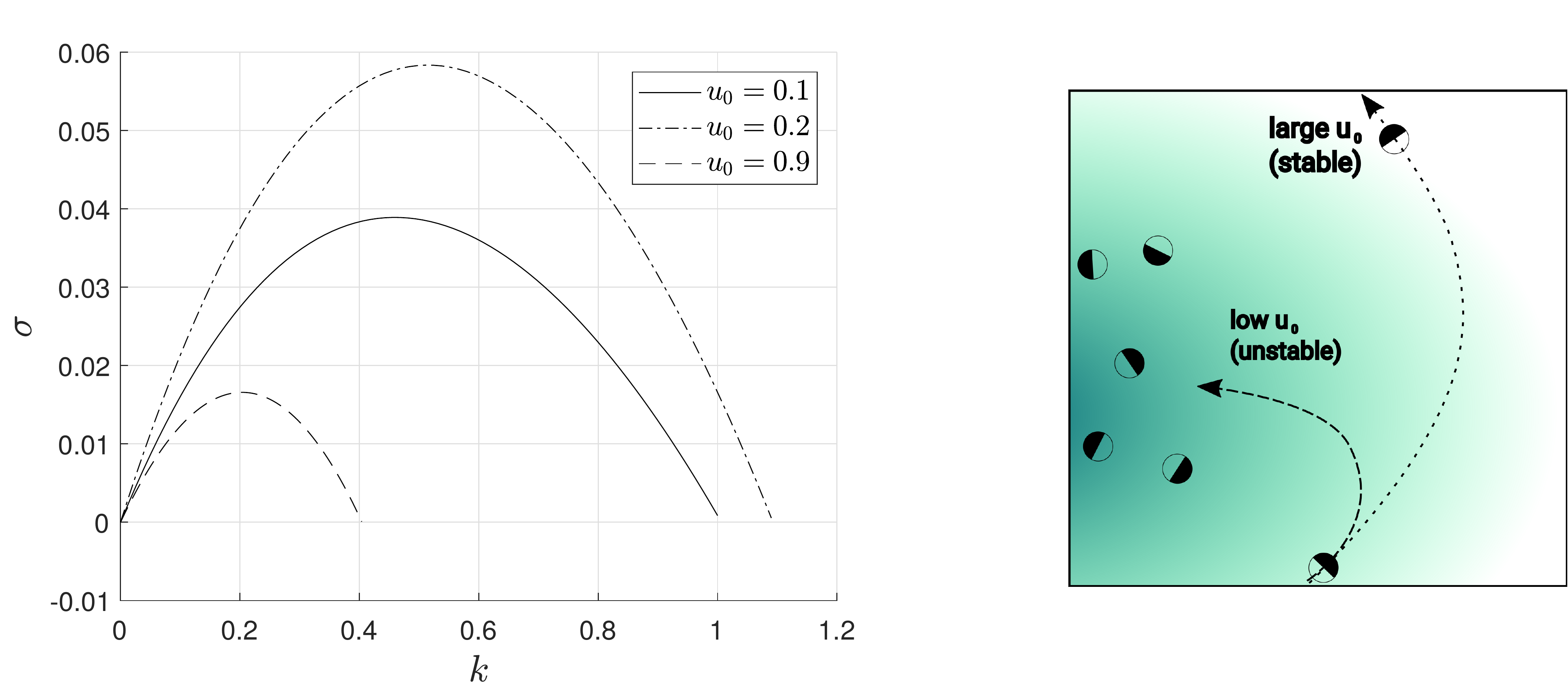}
        \caption{Chemotactic limit. Left: Evolution of the growth rate of the least stable mode as obtained numerically from Eq.\eqref{DispRelChem} in the chemotactic limit ($\xi_r/\varphi =0.6$, $\xi_t=-0.375$) for different values of $u_0$ with $\beta=2\pi$ and $d_x=0.05$. Right: Schematic representation of the chemotactic mechanism and the stabilizing effect of self-propulsion.} \label{xi_inst}
\end{figure}
The chemotactic limit of the instability corresponds to suspensions of particles with positive mobility contrast (${M^->0}$), that provides them with the ability to perform autochemotaxis ($\xi_r>0$). 
In order to rule out the destabilizing effect of phoretic attraction discussed in Section~\ref{phoretic_inst} and to isolate the role of chemical reorientation, we consider particles with positive average mobility  ($M^+ > 0$), corresponding to phoretic repulsion ($\xi_t<0$).

In contrast with the phoretic limit, the physical mechanism that promotes the instability relies on self-propulsion: an excess of chemoattractant produced in regions of higher concentration of colloids induces other particles to reorient and to swim towards it.  
As new swimmers approach this region, they release more chemical and raise its concentration level even more: an instability develops through this positive feedback loop (see Fig.~\ref{xi_inst}, right). This mechanism, which results in an effective attraction between particles, was also studied in Refs.~\cite{Saha2014, Liebchen2017,Lushi2018}.  

Chemical interactions therefore polarize the swimmers, which in turn exploit their self-propulsion capacity to amplify perturbations of the particle concentration, $\Phi$.
This can be observed in Eq.~\eqref{1stEq}, where a forcing of self-propulsion $u_0$ on the particle concentration $\Tilde{\Phi}$ is observed when the polarization $\Tilde{\textbf{n}}$ has a non-zero component along the solute gradient (i.e. perpendicular to the wave-front, $\tilde{\textbf{n}}\cdot\textbf{k} \neq 0$). 
In contrast with the phoretic limit where they are driven solely by the local concentration gradient, fluxes leading to accumulation of particles are here enforced by self-propulsion; 
as a result, long-wavelength modes may still be unstable even though they correspond to weak chemical gradients.

To determine the instability criterion, Eq.~\eqref{DispRelChem} is expanded in the small-$k$ limit, where the effect of the phoretic drift can be neglected , i.e. $\sigma = \sigma_1k+\sigma_2k^2+...$. Retaining only leading-order $O(k^0)$-terms yields  
  \begin{eqnarray}
    \frac{u_0\beta\varphi}{4\pi\xi_r} = 1-\frac{\sigma_1}{u_0}\tan^{-1}\frac{u_0}{\sigma_1} \cdot  \label{eqSlope_xi}
   \end{eqnarray}   
 Instability is obtained for $\sigma_1>0$ which imposes
   \begin{eqnarray}
    0 < \frac{u_0}{(\xi_r/\varphi)} < \frac{4\pi}{\beta} \cdot \label{inst_cond_xi}
   \end{eqnarray}
For a given relaxation rate of the chemoattractant ($\beta$) the existence of the chemotactic instability imposes a maximum ratio between the velocity at which a particle self-propel ($u_0$) and the rate at which it rotates into an external chemical gradient $(\xi_r/\varphi)$. 

In agreement with the numerical solution of the full dispersion relation, Eq.~\eqref{DispRelChem} reported in Fig.~\ref{xi_inst}, Eq.~\eqref{inst_cond_xi} reveals the dual role and impact of self-propulsion on the instability. Swimming is indeed a necessary ingredient as reoriented (polarized) particles need to actively move toward solute-rich regions. However, large swimming speeds can suppress the instability if particles swim past localized regions of higher concentration before reorienting fully toward it. In that case, the trajectory of each swimmer is weakly curved under the effect of the attractive phoretic field, but it eventually escapes, as depicted in Fig.~\ref{xi_inst} (right). The instability criterion can also be reformulated as $\tau_r<\tau_t$, with $\tau_r$ the time scale associated with the particle rotation in response to the chemical gradient ($\tau_r \sim \varphi/(\xi_r k)$)  and $\tau_t$ that associated with swimming over the characteristic length of the mode considered ($\tau_t \sim 1/(u_0 k)$). We thus note that in the dilute limit, $\varphi \ll 1$, the stabilizing effect of $u_0$ is relevant for particles with small mobility imbalance $M^-$, namely when $\xi_r \ll 1$.

%
%

%
%
%
\subsection{Anti-chemotactic limit}  \label{delay_inst}
\begin{figure}
        \includegraphics[scale= 0.4]{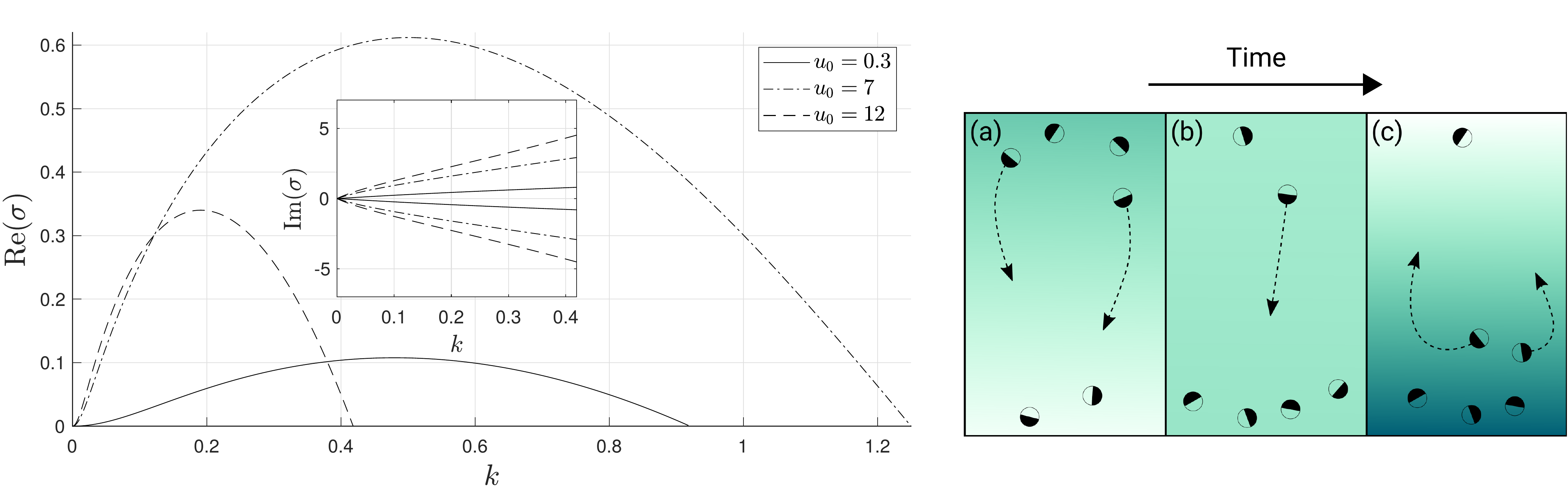}
        \caption{Antichemotactic limit. Left: Evolution of the growth rate and frequency (inset) of unstable modes of wave number $k$ as obtained from the chemical dispersion relation \eqref{DispRelChem} in the anti-chemotactic (delay) limit with $\xi_r/\varphi = -3$, $\xi_t=-0.375$, $\beta=0.628$, $d_x=0.05$. Right: Illustration of the anti-chemotactic instability mechanism: (a) Particles orient and  swim towards regions of low chemical concentration; (b) Particles have reached the region of low solute content and start releasing solute there, but the solute concentration takes time to equilibrate leading to more particles aggregation; (c) In response to the increased particle accumulation, the chemical concentration overshoots (in comparison with (a)), and particles rotate and swim away from the solute-rich region.  } \label{fig:delay_inst}
\end{figure}

The anti-chemotactic limit corresponds to suspensions of particles with negative mobility contrast ($M^-<0$), meaning that they perform negative chemotaxis ($\xi_r<0$) and with positive average mobility ($M^+>0$), corresponding to phoretic repulsion ($\xi_t<0$).
The interplay of negative chemical reorientation and self-propulsion leads to the migration of swimmers toward regions of low solute concentration. The colloids considered here are net chemical sources and thus eventually raise the local chemical concentration around them ($A^+>0$), thereby canceling and reversing the solute gradient that attracted them in the first place, and generating a joint oscillation dynamics of the particle and solute concentration. However, if the solute production is slow enough, a delay will be observed between the particles' accumulation at a given location and the resulting growth of chemical concentration, which will cause particles to escape away. If this delay is large enough, a concentration overshoot will be observed from one period to the next, triggering unstable oscillatory modes \cite{Liebchen2015}. An illustration of the mechanism of this delay instability is provided in Fig.~\ref{fig:delay_inst} (right).

In contrast with the two mechanisms discussed previously in the phoretic and chemotactic limits, the unsteady nature of chemical diffusion is essential here (i.e. $\partial C/\partial t\neq 0$): a quasi-steady assumption would indeed enslave the solute content to particle concentration $\Phi$ preventing its overshoot, and the oscillatory modes discussed above would simply be damped out by particle diffusion.

The dispersion relation, Eq.~\eqref{DispRelChem}, is solved numerically in the anti-chemotactic limit and the results for the unstable modes are reported in Fig~\ref{fig:delay_inst} (left). 
We observe once again that self-propulsion has a dual role: it is necessary for the instability to develop as particles need to be able to swim away from solute-rich regions, and for low $u_0$, the instability is promoted by an increase in swimming velocity (a larger range of wave numbers become unstable and the growth rates are increased). However, the trend is reversed when $u_0$ is too large and self-propulsion tends to suppress the instability, for the same reason as was detailed in the chemotactic limit: particles swimming faster than they reorient are not able to polarize and converge to solute-depleted regions.

In the long-wave limit, the dispersion relation takes the same form as the chemotactic limit, Eq.~\eqref{eqSlope_xi}, and
\begin{equation}
    \frac{\sigma_1}{u_0}\tan^{-1}\frac{u_0}{\sigma_1} = 1-\frac{u_0\beta\varphi}{4\pi\xi_r} \label{eqSlope_xi2}
   \end{equation}   
is a real and positive number. This in turns imposes $\sigma_1$ to be purely imaginary (i.e. the growth rate is at least $O(k^2)$ for small $k$) and its frequency to be greater than $u_0$, which is consistent with $\mbox{Im}(\sigma_1)$ increasing with $u_0$ (see Fig.~\ref{fig:delay_inst}). Furthermore, the mode frequency varies linearly with $k$, as expected from physical argument: the period of oscillation is proportional to the time necessary for the particles to swim from regions of high and low concentrations that are typically distant by half a wavelength so that $\mbox{Im}(\sigma)\sim u_0 k$. Consequently, in the long-wavelength limit, the dynamics become very slow and the evolution of the phoretic field approaches the quasi-steady regime, instantaneously determined by the density $\Phi$, yielding Re$(\sigma_1)=0$. 
%
%
\subsection{Summary of the destabilizing mechanisms}  \label{sec:summary_inst}
%
We have discussed three different destabilizing mechanisms of isotropic and uniform suspensions promoted solely by chemical interactions and leading to particle aggregation. These were characterized by increasing
complexity as: 
\begin{enumerate}[(i)]
\item{The phoretic limit, based solely on the translational drift of particles along attractive chemical gradients; }
\item{The chemotactic limit, which is based on
the cooperation between self-propulsion and chemical reorientation;}
\item{ The anti-chemotactic mechanism, which requires additionally the phoretic field to be unsteady. }
\end{enumerate}
Hydrodynamic interactions are absent in this linear limit due to the spherical shape of the colloids which do not respond to shear-alignment as for slender rods or bacteria. They may however play a role in the nonlinear evolution of the perturbation, and in the following, we therefore turn our attention to the effect of hydrodynamic interactions on the long-term  dynamics.

%
%
%
%
\section{Hydrodynamically-induced disorder} \label{sec:num_sim}
In the dilute limit, hydrodynamic interactions between swimming particles are dictated by their dominant hydrodynamic signatures, i.e. their stresslet. As noted in Eqs.~\eqref{sigma_s}, \eqref{Ind_stresslet2} and \eqref{nondim_stress_int}, the particle stresslet intensity is proportional to the mobility contrast $M^-$ regardless of its origin (i.e the particle's own activity or a background chemical gradient) and is therefore proportional to the chemically-induced reorientation, $\xi_r$.   
Consequently the stresslet intensities $\alpha_s$ and $\alpha_i$ and hydrodynamic signature of the Janus particles are negligible in the phoretic limit ($\xi_r \rightarrow 0$): in that case, there is therefore no hydrodynamic effect since there is no flow generated in the dilute limit. 

We therefore focus our attention in the following on the chemotactic and anti-chemotactic limits of the chemical instability and analyse in those cases how the emergence of hydrodynamic flow in the nonlinear regime influences the saturated dynamics of the system. 
We finally remark that, unlike for microorganisms,  the sign of the self-induced stresslet $\alpha_s$ of Janus particles is tied to the attractive or repulsive nature of their chemical interactions.
Specifically, in the chemotactic and anti-chemotactic limits particles are pusher ($\alpha_s<0$) and puller ($\alpha_s>0$) swimmers, respectively (see Eq.~\eqref{nondim_stress_int}). 
%
%
%
\subsection{Numerical method} 
We investigate the interplay of hydrodynamic interactions and chemical signaling in the non-linear regime in a two-dimensional limit by solving numerically Eqs.~\eqref{EvolEqPsi_dim}, \eqref{conteq}, \eqref{momeq} and \eqref{conceq} in a square
periodic domain; the particle and solute distributions are therefore assumed invariant in the third ($z$) direction, and particles are oriented within the $(x,y)$-plane. Note that the latter is physically relevant when significant concentration gradients are confined within that plane as chemical reorientation tends to align the particles' axis with or against such gradients. 
The $z$-invariance assumption significantly reduces the computational cost and is often used to solve numerically similar kinetic models \cite{Saintillan2008,Lushi2012,Lushi2018}. Full three-dimensional simulations are performed for suspensions of elongated bacteria undergoing a purely hydrodynamic instability with \cite{Saint2019_big_num} and without \cite{Saint2011_shear_eff} confinement. It emerges that the observed 3D patterns closely resemble the ones observed in 2D simulations, suggesting that the present approach provides correct qualitative predictions of the suspension's dynamics. We note that, to the best of our knowledge, three-dimensional simulations of kinetic models for dilute suspensions of autochemotactic microswimmers have not been discussed yet in the literature.      

Simulations are performed using a 128-by-128 grid in the physical space $\textbf{x}=[x,y]$ with a non-dimensional box size of $L=30$ and using 32 points to discretize the orientational dynamics of the particle on the plane $\textbf{p}=[\cos\theta,\sin\theta]$.
The system is solved using a spectral method: Stokes equations are solved in Fourier space and the nonlinear terms in Eqs.~\eqref{EvolEqPsi_dim} and \eqref{conceq} are computed in physical space performing a grid augmentation to avoid aliasing. A $4^\textrm{th}$-order Runge-Kutta scheme is used for time-marching.
In all simulations, the translational and rotational diffusion are set to $d_x=d_r=0.025$.

Initially, a small perturbation in particle distribution (in space and orientation) is  added to the uniform and isotropic state a perturbation of the form $\delta\Psi_{0} = \sum_{j} \epsilon_j(\theta)\cos( \textbf{k}_j\cdot \textbf{x} + \theta^*_j )  $ with $1\leq j\leq 15$, where $\epsilon_j(\theta)$ is a third-order polynomial in $\cos\theta$ and $\sin\theta$ with random $O(10^{-3})$ coefficients and $\theta^*_j$ is a random phase. The phoretic field is initiated with a uniform distribution $C_0=2\pi/\beta$.

\subsection{Chemotactic limit} \label{sec:chem_numeric}

\begin{figure}[htb] 
  \centering
        \includegraphics[scale= 0.55,angle=0]{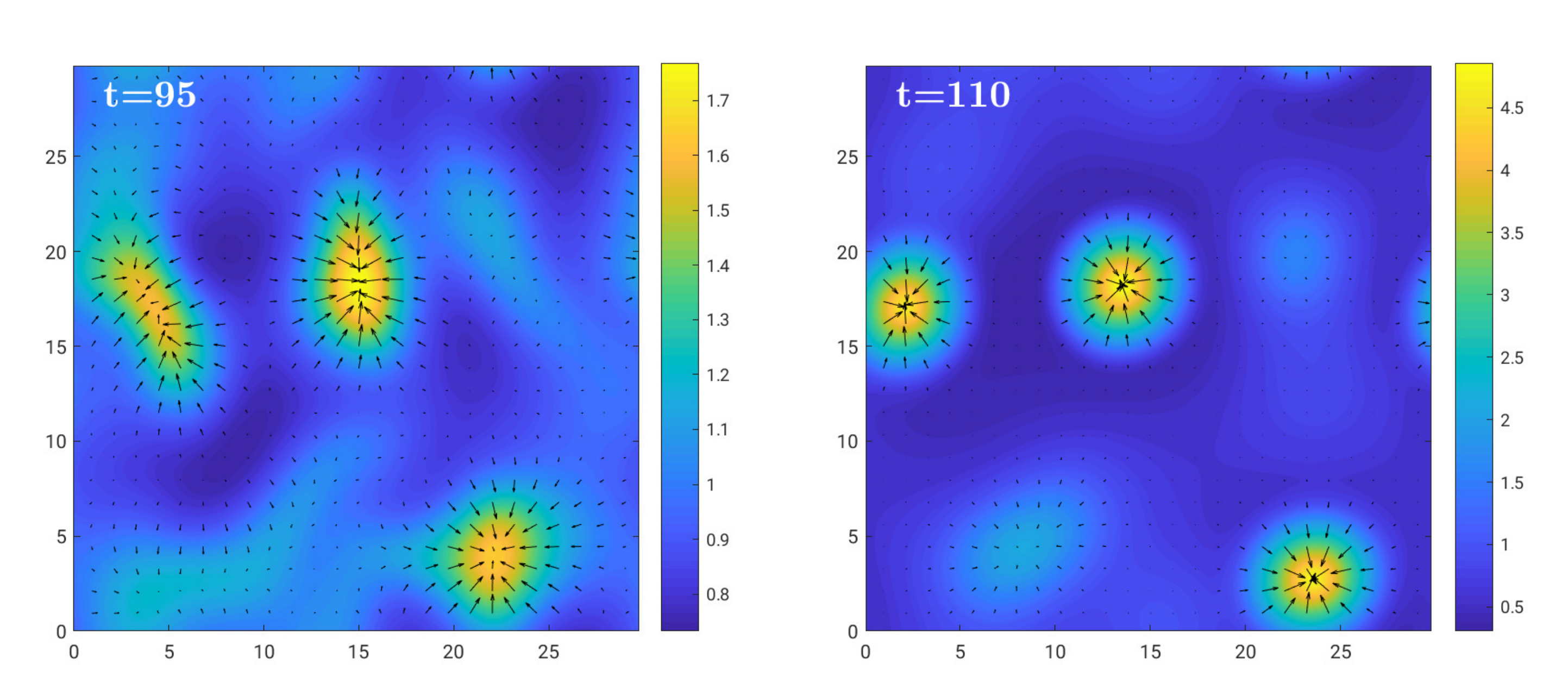}
         \caption{Emergence of nonlinear dynamics in the chemotactic limit ($\xi_r/\varphi = 1$, $\xi_t = -0.375$, $u_0=0.5$, $\beta=2\pi$). Particle density $\Phi$ (color) and mean direction field $\textbf{n}$ (arrows) forming \textit{asters}. Left: with hydrodynamic interactions ($\alpha_s=   -3.1169$, $\alpha_i=0.3142$). Right: without hydrodynamic interactions ($\alpha_s=\alpha_i=0$).
         } \label{fig:asters}
\end{figure}

\begin{figure}[htb] 
  \centering
        \includegraphics[scale= 0.61,angle=0]{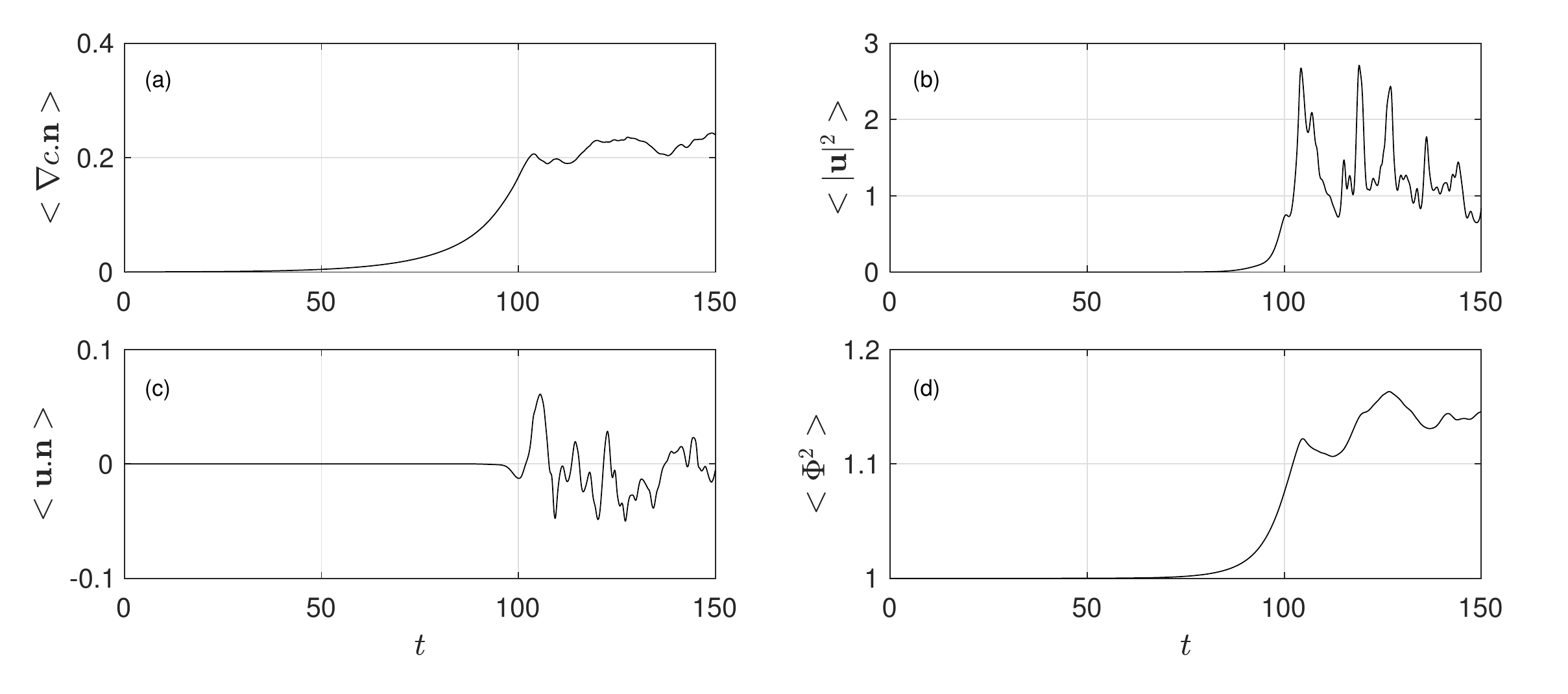}
         \caption{Time series of relevant quantities in the chemotactic limit ($\xi_r/\varphi = 1$, $\xi_t = -0.375$, $u_0=0.5$, $\beta=2\pi$); $<\bullet>$ indicates spatial averages. (a) Spatial correlation of the chemical gradient and the mean director, (b) variance of the flow velocity, (c) spatial correlation of the flow velocity and the mean director, (d) variance of the particle density.
         } \label{fig:contour_n_u_GradCn}
\end{figure}

Chemical reorientation and alignment of the Janus particles with  local chemical gradient  drives the system away from the isotropic state to a configuration with a net polarization (Fig.~\ref{fig:asters}), also referred to as \textit{asters} in \cite{Saha2014}. 
The spatial correlation of the chemical gradient and the mean director field is computed, $<\nabla C \cdot \textbf{n}>$, to quantify the effect of chemical signaling on the particles orientation. 
The evolution of the flow intensity is also evaluated through the variance of the flow velocity, $<|\textbf{u}|^2>$, as a direct measure of the strength of the hydrodynamic interactions between particles, see Fig.~\ref{fig:contour_n_u_GradCn}. 
During the initial exponential growth of the perturbation, which lasts up to $t\sim 90$, $<|\textbf{u}|^2>$ and thus hydrodynamic interactions remain negligible while $<\nabla C \cdot \textbf{n}>$ grows exponentially, confirming that the dominant interactions are chemically-mediated during the astering process. 

In a second phase, the emergence of a local polar order allows for the cumulative hydrodynamic effect of many particles, resulting in the emergence of a large-scale fluid flow. 
The inward flux of particles due to convection is $-\nabla_x \cdot (\Phi\textbf{u})$ which reduces to $- \textbf{u}\cdot \nabla_x \Phi$ for an incompressible fluid 
meaning that advection can raise the particle concentration at a given location only if there exists a neighboring location where $\Phi$ is already higher. 
The astering process consists in the accumulation of particles towards locations which are local maxima of $\Phi$ (Fig.~\ref{fig:asters}, right) 
hence such process can not be enforced by the presence of the fluid flow. Consequently, advection of both particles and solute by the particle-generated flows inevitably stretches and breaks the patterns generated by chemical interactions.  
We remark that if the fluid was compressible or if the inertia of the colloids was not negligible (i.e. possibility of cross-streamline migration) the effect of hydrodynamic transport could in principle enforce the astering process.  

\begin{figure}[htb] 
  \centering
        \includegraphics[scale= 0.6,angle=0]{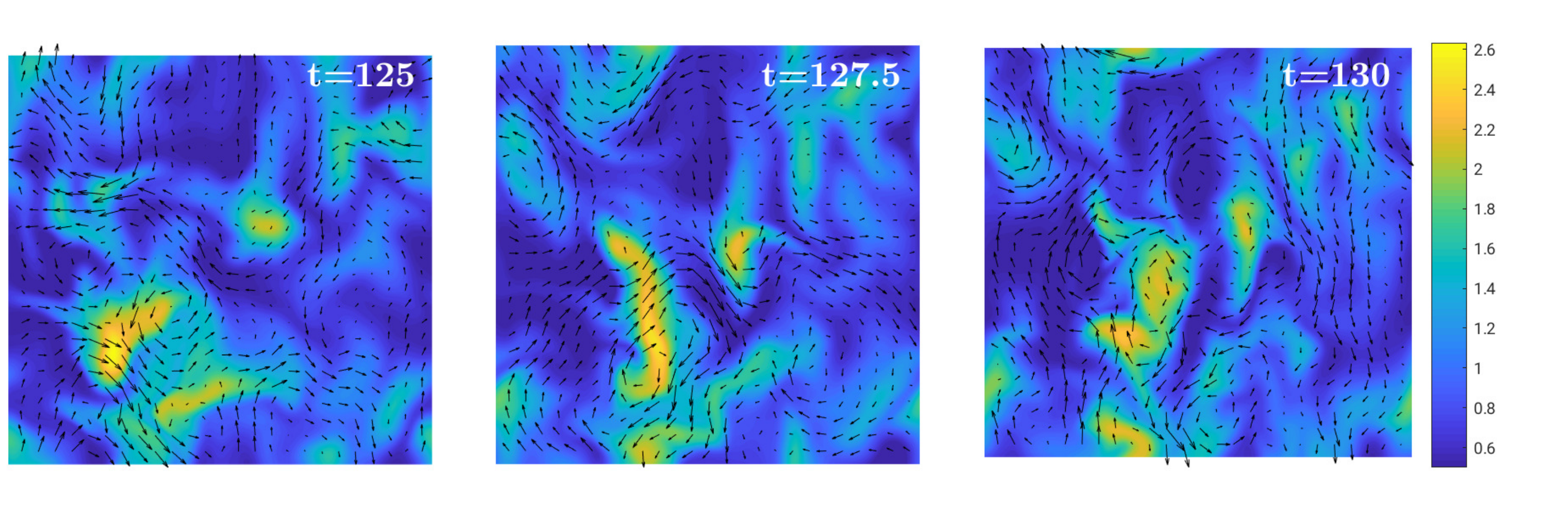}
        \caption{Nonlinear evolution of the particle density $\Phi$ (color) and fluid velocity $\textbf{u}$ (arrows) in the chemotactic limit ($\xi_r/\varphi = 1$, $\xi_t = -0.375$, $u_0=0.5$, $\beta=2\pi$). The entire domain is depicted as in Fig.~\ref{fig:asters}.} \label{fig:timelaps}
\end{figure}

The spatial distribution of particles and chemoattractant is therefore rearranged by such flow and 
particles respond by turning to the new direction of the local chemical gradient. 
The underlying chemotactic mechanism still promotes aggregation of particles but now the resulting aggregates are dynamic because of hydrodynamic transport. 
The system approaches a chaotic attractor where the underlying dynamics are cyclical, under the conflicting influence of chemotactic aggregation and hydrodynamic stirring by the flow driven by the locally-polarized suspension: 
\begin{enumerate}[(i)]
\item{Denser regions form due to chemotaxis and more particles align in the resulting chemical gradient;}
\item{This growing local polar order enhances the generated flow field (see peaks of $<|\textbf{u}|^2>$, Fig.~\ref{fig:contour_n_u_GradCn});}
\item{Regions of larger particle (and solute) density are stretched and mixed under the effect of hydrodynamics and diffusion (Fig.~\ref{fig:timelaps});} 
\item{The polar order decreases resulting in a weaker  flow;}
\item{Particles aggregate again in the quiescent flow and the cycle starts over again.}
\end{enumerate}  
Such cyclical dynamics translates into an actual stirring effect and prevent the asters from collapsing, therefore the variance of the particle density, $<\Phi^2>$, saturates not far from its base-state value $<\Phi_0^2>=1$ (Fig.~\ref{fig:contour_n_u_GradCn}).

Our results show that the role of hydrodynamics is crucial even for spherical particles that do not lead to any purely hydrodynamic instability. 
Under these circumstances, the correlation between the flow field and the local mean orientation of the particles, $<\textbf{u}\cdot\textbf{n}>$, averages to zero over time (Fig.~\ref{fig:contour_n_u_GradCn})
because geometrically-isotropic swimmers do not experience shear-alignment (unlike rod-like pushers or disk-like puller \cite{Saintillan2008,Nejad2019}) and polarize only in the far-field chemical signature of other particles.

\begin{figure}[htb] 
  \centering
        \includegraphics[scale= 0.55,angle=0]{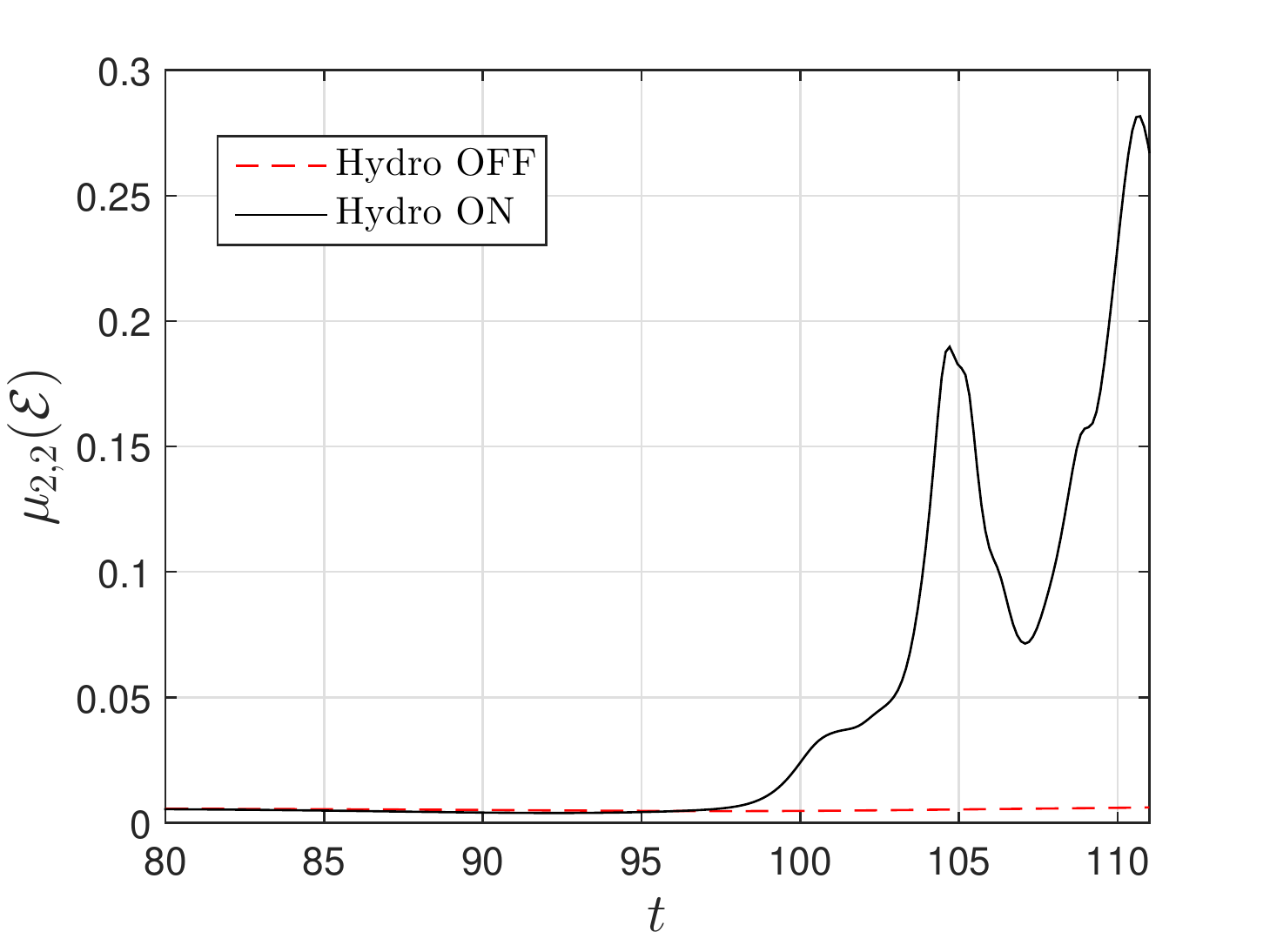}
         \caption{
         Time series of the second central moment of the normalized energy density $ \mu_{2,2} (\mathcal{E})$ in the chemotactic limit ($\xi_r/\varphi = 1$, $\xi_t = -0.375$, $u_0=0.5$, $\beta=2\pi$) with and without hydrodynamic interactions.
         } \label{fig:Energy_spectrums}
\end{figure}
\begin{figure}[htb] 
  \centering
        \includegraphics[scale= 0.52,angle=0]{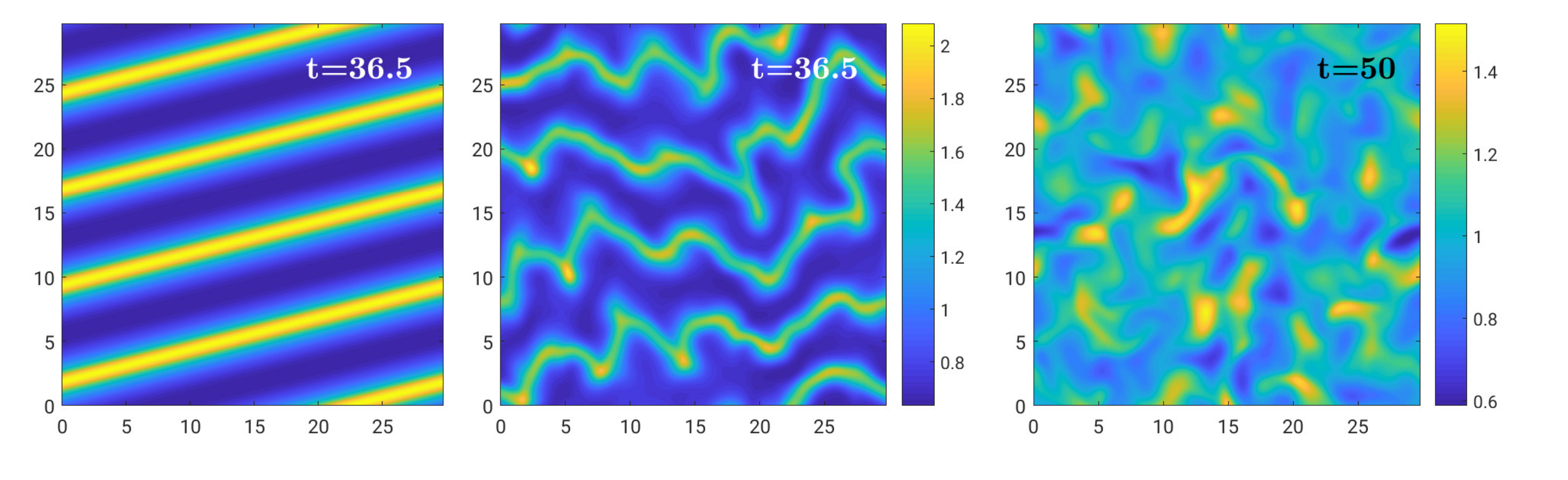}
         \caption{Contour plots of the particle density $\Phi$ in the anti-chemotactic limit ($\xi_r/\varphi = -3$, $\xi_t=-0.5$, $u_0=2$, $\beta=2\pi/3$). Left: without hydrodynamic interactions ($\alpha_s$, $\alpha_i=0$). Center and right: with hydrodynamic interactions ($\alpha_s=28.05$, $\alpha_i=-0.94$).} \label{fig:Delay_inst}
\end{figure}
\begin{figure}[htb]
\centering                             %
 \includegraphics[scale= 0.55,angle=0]{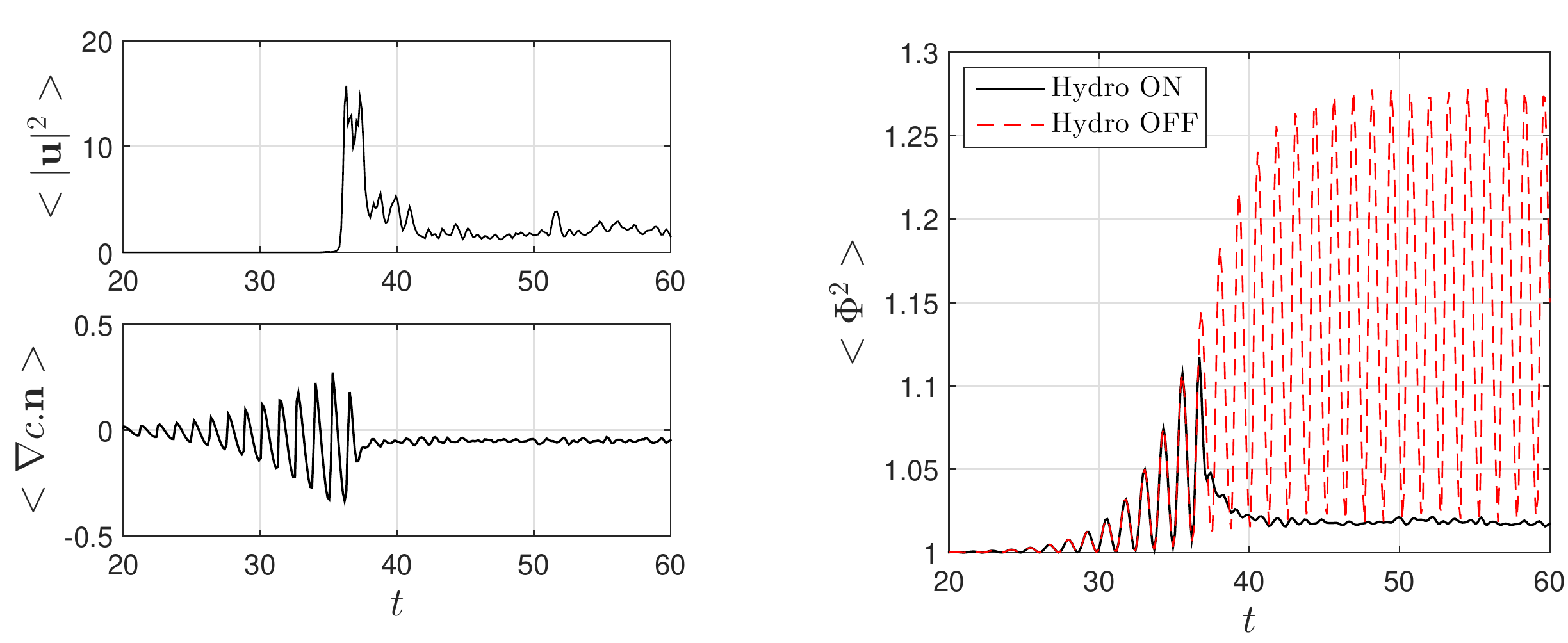}
\caption{Time series of relevant quantities in the antichemotactic limit ($\xi_r/\varphi = -3$, $\xi_t=-0.5$, $u_0=2$, $\beta=2\pi/3$). Left: variance of the flow velocity (top) and spatial correlation of the chemical gradient and the mean director (bottom).  Right: Variance of the particle density $<\Phi^2>$ with and without hydrodynamic interactions.
         } \label{fig:Fi_mean_delay}
\end{figure}
To improve our understanding on the relative importance of hydrodynamic and chemical coupling, the same simulation is performed with and without hydrodynamic interactions. The latter is obtained by artificially setting $\alpha_s=\alpha_i=0$; in the present dilute model, such particles do not generate any flow forcing so $\mathbf{u}=0$. Note that this artificial situation can not be reached through a specific choice of activity and mobility distribution for the Janus phoretic particles; indeed, we demonstrated in Section~\ref{subsec:non_dim} that the stresslet intensities are not independent parameters of the problem. 
Asters form whose typical size and circular shape is determined by the wavelength of the most unstable mode. 
Due to the absence of advective flow, these structures are neither stretched nor broken, maintaining a regular shape in time (Fig.~\ref{fig:asters}, right).
The properties of the energy spectrum of the particle density  
reflects the fundamental difference between the  system's attractors, with and without hydrodynamics. 
Precisely, we are interested in the normalized energy spectrum
$\mathcal{E}(\textbf{k},t) \coloneqq |\Hat{\delta\Phi}|^2 /  \int |\Hat{\delta\Phi}|^2 \textrm{d}^2\textbf{k}$,
where $\Hat{\delta\Phi}$ is the two-dimensional Fourier transform of the perturbation of the particle density $\delta\Phi = \Phi - \Phi_0$ and $\textbf{k} = (k_x,k_y)$ is the two-dimensional wave vector. 
$\mathcal{E}(\textbf{k},t)$ can be interpreted as the probability density  
of finding a particle density wave with wave vector $\textbf{k}$ at time $t$. 
Equivalently, the ratio $\mathcal{E}(\textbf{k}_1,t)/\mathcal{E}(\textbf{k}_2,t)$ can be seen as the most likely relative amplitude of two particle density waves with wave vectors $\textbf{k}_1$ and $\textbf{k}_2$, at time $t$.   
We then compute the second central moment of $\mathcal{E}$ on the Fourier plane, defined as
\begin{eqnarray}
   \mu_{2,2} (\mathcal{E}) = \int (k_x-\mathbb{E}[k_x])^2(k_y-\mathbb{E}[k_y])^2 \mathcal{E}(\mathbf{k}) \textrm{d}^2\textbf{k} ,
\end{eqnarray} 
where $\mathbb{E}[k_{x,y}] = \int k_{x,y} \mathcal{E}(\mathbf{k})\textrm{d}^2\textbf{k}$.
The vector $\textbf{k}_e = (\mathbb{E}[k_x],\mathbb{E}[k_y])$ is the expected dominant wave vector while the magnitude of $\mu_{2,2} (\mathcal{E})$ represents an intrinsic measure of how scattered the energy density is on the $(k_x,k_y)$-plane.       
Without hydrodynamics, the energy of the spatial signal $\delta\Phi$ concentrates around those wave numbers corresponding to the chemically-unstable modes,  
resulting in a smaller value of $ \mu_{2,2} (\mathcal{E})$. 
In contrast, the presence of an induced flow field continuously depletes energy from those modes, which is injected at higher wavenumbers that were chemically-stable and where particle diffusion dominates. 
A broader power spectrum for $\delta\Phi$ (i.e. larger variety of length scales in the particle distribution) corresponds to a larger value of $ \mu_{2,2} (\mathcal{E})$, as can be seen in Fig.~\ref{fig:Energy_spectrums}.  
\subsection{Anti-chemotactic limit} \label{sec:antichem_numeric}
Hydrodynamic interactions and flow-induced stirring  also play a major role in the dynamics of the system in the anti-chemotactic limit, where particles are pullers, namely $\alpha_s>0$, see Eq.~\eqref{nondim_stress_int}. To analyze this, the nonlinear dynamic equations are solved numerically in the anti-chemotactic limit, with and without hydrodynamics. 

Without hydrodynamic interactions (i.e. setting artificially $\alpha_s=\alpha_i=0$), a similar dynamics to that obtained by means of particle-based simulations in Ref.~\cite{Liebchen2015} is observed: After an initial transient, traveling density waves appear, corresponding to bands of particles escaping their own chemical footprint, Fig.~\ref{fig:Delay_inst} (left). 
The evolution in time of the particle density variance is also reported in Fig.~\ref{fig:Fi_mean_delay} (right), showing that the solution without hydrodynamics is characterized by regular oscillations which result from the interference between density waves traveling in different directions. 
The amplitude of such oscillations initially grows exponentially before saturation is reached due to particle diffusion.  

When hydrodynamic interactions are properly accounted for (i.e. by setting $\alpha_s$ and $\alpha_i$ to their values determined by the particles' chemical properties), we observe an initial transient during which the generated flow disturbance is very weak (Fig.~\ref{fig:Fi_mean_delay}, top left) and the solution is indistinguishable from the one obtained without hydrodynamic effects. As the amplitude of the oscillating density waves grows, the polarization of the swimmers is enhanced by the resulting chemical gradient and so is the amplitude of oscillation of the spatial correlation $<\nabla C \cdot \textbf{n}>$  (Fig.~\ref{fig:Fi_mean_delay}, bottom left). 
Strong particles' polarization results in cumulative flow forcing that distorts the otherwise regular wave fronts (see Fig.~\ref{fig:Delay_inst}, center) eventually inducing a net unsteady flow that stirs and mixes the unstable spatio-temporal patterns driven by the chemically-induced particle reorientation (Fig.~\ref{fig:Delay_inst}, right). Due to such hydrodynamically-induced disorder the oscillating nature of the underlying destabilizing mechanism is no longer evident: $<\nabla C \cdot \textbf{n}>$ eventually saturates at a negative value meaning that particles on average points away from regions of high solute concentration, as expected for antichemotactic swimmers.

As can be observed by the time series of $<\Phi^2>$ in Fig.~\ref{fig:Fi_mean_delay}, the stirring effect of the self-generated flow within the suspension significantly limits particle accumulation as in the chemotactic limit. 
Remarkably, we find that hydrodynamic interactions induce a similar stirring effect in suspensions of pushers and pullers spherical Janus particles unlike for suspensions of elongated microorganisms~\cite{Lushi2018,Lushi2012}. 
%

 %
%
%
%
\section{Conclusions}
\label{sec:conclusions}
In summary, the present work provides a novel insight on the direct link between the microscopic properties of spherical Janus colloids and the resulting macroscopic stability and dynamics of dilute suspensions of such colloids. 
To achieve this, a kinetic model was used to account for the particle self-propulsion as well as their chemical and hydrodynamic interactions mediated by their environment through mean ambient fields. As all these characteristics are set directly by the fundamental distribution properties of the particles' mobility and activity (namely their mean value and contrast between the two sides of the Janus colloids),  chemically-mediated interactions within the suspension and the individual self-propulsion velocity of the particles are intricately related to the hydrodynamic disturbance introduced by the swimmers, and reciprocally. 
By accounting for such a link the present approach therefore allows us to investigate the reciprocal interplay of these three components (self-propulsion, chemical and hydrodynamic coupling). 

Within the dilute limit, isotropic suspensions of spherical particles are shown to be unstable to small perturbations and different regimes of instability are identified depending on the distribution of phoretic mobility at the particles' surface (which in turn influences their ability to drift and reorient within external chemical gradients and thus their chemotactic or anti-chemotactic behavior). Because hydrodynamic coupling (and in particular shear-alignment) is negligible for spherical colloids in this linear limit, the emergence of instabilities is purely due to the chemical signaling and coupling of the different particles. Yet, local coupling of chemical and hydrodynamic processes (i.e. at the particle level) plays a critical role in the development of such instabilities as it directly impacts the self-propulsion velocity of the colloids.

The magnitude of self-propulsion velocity 
is indeed shown to critically affect the linear stability of the system and the wave selection mechanism of the most unstable perturbations, but it does so in different ways depending on the properties of the surface of the colloids and the ensuing dominant instability regime. 
For suspensions of particles with uniform mobility which are phoretically attracted to each other (i.e. phoretic limit), the presence of self-propulsion has a purely stabilizing effect and  introduces a wave selective mechanism in the linear regime. %
On the other hand, if the mobility contrast of the swimmers is non-negligible, particles are able to reorient along or against gradients of ambient solute concentration (i.e. they are chemotactic or anti-chemotactic). In these cases, self-propulsion is a necessary ingredient for the instability to exist as it allows reorienting particles to migrate along or against such chemical gradient. 
Yet, interestingly, for small mobility contrast (i.e. slow reorientation of the colloids), self-propulsion also has a dual effect as increasing values of $u_0$ stabilize the suspension. 

The magnitude of the self-propulsion velocity plays therefore a key role in setting the main features of the macroscopic collective dynamics in the linear regime (e.g. dominant length scale and growth rate). This velocity is directly controlled by the activity contrast of the colloid (i.e. its ability to generate gradients between its two faces), which identifies a route for direct design control of the emergence of such instabilities through the activity distribution. 
For example, suspensions of particles with either uniform or weakly-non-uniform mobility could essentially be completely stabilized by increasing the self-propulsion velocity or the activity contrast of the particles. 

The strength of the far-field hydrodynamic footprint generated by each particle (stresslet), and its sign (which sets its pusher or puller characteristic) is also directly proportional to the mobility contrast. Yet, the classical hydrodynamic instability observed for pushers (e.g. bacterial suspensions~\cite{Saintillan2008}) is not observed here for spherical swimmers. However, this does not mean that hydrodynamic interactions do not play any role in the suspension dynamics. In fact, our numerical results on the non-linear suspension dynamics resulting from the saturation of the initial instabilities, show that the long-ranged chemically-induced polarization of particles induces coherent hydrodynamic forcing on the fluid. Consequently, we observe the emergence of a strong hydrodynamic field for positive as well as negative autochemotactic swimmers, which in turn corresponds here to pusher and puller swimmers. For both kind of swimmers, such generated fluid flow is responsible for a stirring effect which limits accumulation of particles and mixes and suppresses regularly patterned particle distribution (and solute concentration) promoted by chemical signaling. 
This process is dynamic: once such hydrodynamically-induced disorder reduces the long-range chemically-induced polarization, the coherence of flow forcing by the particles breaks down which in turn reduces the hydrodynamic flow field and its mixing action. The resulting chaotic dynamics are fundamentally cyclical and characterized by sharp peaks in the intensity of the flow field, followed by temporary accumulation of solute and particles under the effect of chemical coupling.

While previous works had already identified different limits of the chemical instabilities, the present work proposes a unique insight onto the reciprocal importance of hydrodynamic and chemical interactions within suspensions of autophoretic swimmers. The relative weight of these interaction routes has recently received much attention in the physics and hydrodynamics communities. Such problem was directly addressed in the recent work of Ref. \cite{Liebchen2019} using a particle-based representation of the suspension, as opposed to the mean-field description employed here. Such approach naturally accounts for steric particle-particle interactions and it is ideal to study relatively crowded suspensions where far-field hydrodynamic effects arguably play a secondary role and can therefore be neglected. By doing so they successfully reproduce a dynamic-cluster phase similar to the one experimentally observed \cite{Theurkauff2012,Ginot2018} within suspensions of particles with estimated uniform mobility (i.e. negligible far-field hydrodynamic signature). 
By focusing on particles with nonuniform mobility (i.e. with nonzero stresslet intensity) and on the dilute limit, we safely neglect steric effects and we capture the combined effects of the generated hydrodynamic and chemical fields. Under this conditions, our numerical results suggest that it is precisely the cooperation between chemical and hydrodynamic couplings which characterizes the long-term dynamics of the suspension, which would be otherwise profoundly different if any one of the two interaction routes was neglected.  

Finally we remark that the present numerical approach was applied here to a quasi-two-dimensional system where concentration and hydrodynamic fields as well as particle distribution are independent of a third dimension, and as a result particle motion occurs within a two-dimensional plane. Yet, the formalism is completely generic and can be directly used to consider the general 3D case. The effect of rigid boundaries as well as particle confinement on their interactions could also  easily 
be included in the present framework, which will be of particular interest to understand the influence of their complex environment on the collective behavior of phoretic suspensions.

\section*{Acknowledgments}
This work was supported by the European Research Council (ERC) under the European Union's
Horizon 2020 research and innovation program (Grant Agreement No. 714027 to S.M.).

%

%
%


\end{document}